\documentclass[sigconf]{acmart}
\usepackage{amsmath}
\usepackage{mathrsfs}
\usepackage{amsmath}
\usepackage{algorithm}
\usepackage{algorithmic}
\usepackage{multirow}
\usepackage{bm}
\usepackage{subfigure}
\usepackage{epstopdf}
\usepackage{flushend}
\usepackage{balance}
\usepackage{flushend}
  
\AtBeginDocument{%
  \providecommand\BibTeX{{%
    \normalfont B\kern-0.5em{\scshape i\kern-0.25em b}\kern-0.8em\TeX}}}

\copyrightyear{2021}
\acmYear{2021}
\setcopyright{iw3c2w3}
\acmConference[WWW '21]{Proceedings of the Web Conference 2021}{April 19--23, 2021}{Ljubljana, Slovenia}
\acmBooktitle{Proceedings of the Web Conference 2021 (WWW '21), April 19--23, 2021,
Ljubljana, Slovenia}
\acmPrice{}
\acmDOI{10.1145/XXXXXX.XXXXXX}
\begin{document}
\title{Task-adaptive Neural Process for User Cold-Start Recommendation}

\author{Xixun Lin}
\affiliation{
  \institution{Institute of Information Engineering, Chinese Academy of Sciences  \\ School of Cyber Security, University of Chinese Academy of Sciences}
  \country{linxixun@iie.ac.cn}
 }

\author{Jia Wu}
\affiliation{
  \institution{Department of Computing, \\ Macquarie University}
 \country{jia.wu@mq.edu.au}
}

\author{Chuan Zhou$^{*}$}
\thanks{$^{*}$Corresponding author.}
\affiliation{
  \institution{Academy of Mathematics and Systems Science, CAS  \\ School of Cyber Security, University of Chinese Academy of Sciences}
  \country{zhouchuan@amss.ac.cn}
}

\author{Shirui Pan}
\affiliation{
  \institution{Faculty of Information Technology, \\ Monash University}
  \country{shirui.pan@monash.edu}
}

\author{Yanan Cao}
\affiliation{
  \institution{Institute of Information Engineering, \\ Chinese Academy of Sciences}
  \country{caoyanan@iie.ac.cn}
 }

\author{Bin Wang}
\affiliation{
  \institution{Xiaomi AI Lab, \\ Xiaomi Inc}
  \country{wangbin11@xiaomi.com}
}

\def\authors{Xixun Lin, Jia Wu, Chuan Zhou, Shirui Pan, Yanan Cao, Bin Wang}
\renewcommand{\shortauthors}{}
\begin{abstract}
User cold-start recommendation is a long-standing challenge for recommender systems due to the fact that only a few interactions of cold-start users can be exploited. Recent studies seek to address this challenge from the perspective of meta learning, and most of them follow a manner of parameter initialization, where the model parameters can be learned by a few steps of gradient updates. While these gradient-based meta-learning models achieve promising performances to some extent, a fundamental problem of them is how to adapt the global knowledge learned from previous tasks for the recommendations of cold-start users more effectively.
\par
In this paper, we develop a novel meta-learning recommender called task-adaptive neural process (TaNP). TaNP is a new member of the neural process family, where making recommendations for each user is associated with a corresponding stochastic process. TaNP directly maps the observed interactions of each user to a predictive distribution, sidestepping some training issues in gradient-based meta-learning models. More importantly, to balance the trade-off between model capacity and adaptation reliability, we introduce a novel task-adaptive mechanism. It enables our model to learn the relevance of different tasks and customize the global knowledge to the task-related decoder parameters for estimating user preferences. We validate TaNP on multiple benchmark datasets in different experimental settings. Empirical results demonstrate that TaNP yields consistent improvements over several state-of-the-art meta-learning recommenders.
\end{abstract}

\ccsdesc[500]{Information systems~Recommender systems}
\ccsdesc[500]{Computing methodologies~Neural networks}

\keywords{User cold-start recommendation, Meta learning, Neural process}

\maketitle

\section{Introduction}
Recommender systems have been successfully applied into a great number of online services for providing precise personalized recommendations~\cite{zhang2019deep}. 
Traditional matrix factorization (MF) models and popular deep learning models are among the most widely used techniques, predicting which items a user will be interested in via learning the low-dimensional representations of users and items~\cite{koren2009matrix,zheng2016neural,He2017NeuralCF,liu2018social,bigi2021}. These models typically work well when adequate user interactions are available, but suffer from cold-start problems. Recommending items to cold-start users who have very sparse interactions, also known as user cold-start recommendation~\cite{bharadhwaj2019meta,lee2019melu,Dong2020MAMOMM}, is one of the major challenges.
\par
To address user cold-start recommendation, early methods~\cite{lops2011content,bianchi2017content} focus on using side-information, $i.e.$, user profiles and item contents, to infer the preferences of cold-start users. Additionally, many hybrid models integrating side-information into collaborative filtering (CF) are also proposed~\cite{kouki2015hyper,wang2015collaborative}. For example, collaborative topic regression~\cite{wang2011collaborative} combines probabilistic topic modeling~\cite{blei2003latent} with MF to enhance model capability of out-of-matrix prediction. However, such informative contents are not always accessible due to the issue of personal privacy~\cite{xin2014controlling}, and manually converting them into the useful features of users and items is a non-trivial job~\cite{das2007google}.
\par
Inspired by the huge progress on few-shot learning~\cite{snell2017prototypical} and meta learning~\cite{finn2017model}, there emerge some promising works~\cite{vartak2017meta,bharadhwaj2019meta,lee2019melu,du2019scenariometa} on solving cold-start problems from the perspective of meta learning, where \emph{making recommendations for one user is regarded as a single task} (detailed in Definition \ref{def1}). In the training phase, they try to derive the global knowledge across different tasks as a strong generalization prior. When a cold-start user comes in the test phase, the personalized recommendation for her/him can be predicted with only a few interacted items are available, but does so by using the global knowledge already learned.
\par
\begin{figure}
\begin{center}
\includegraphics[width=8.5cm,height=4.5cm]{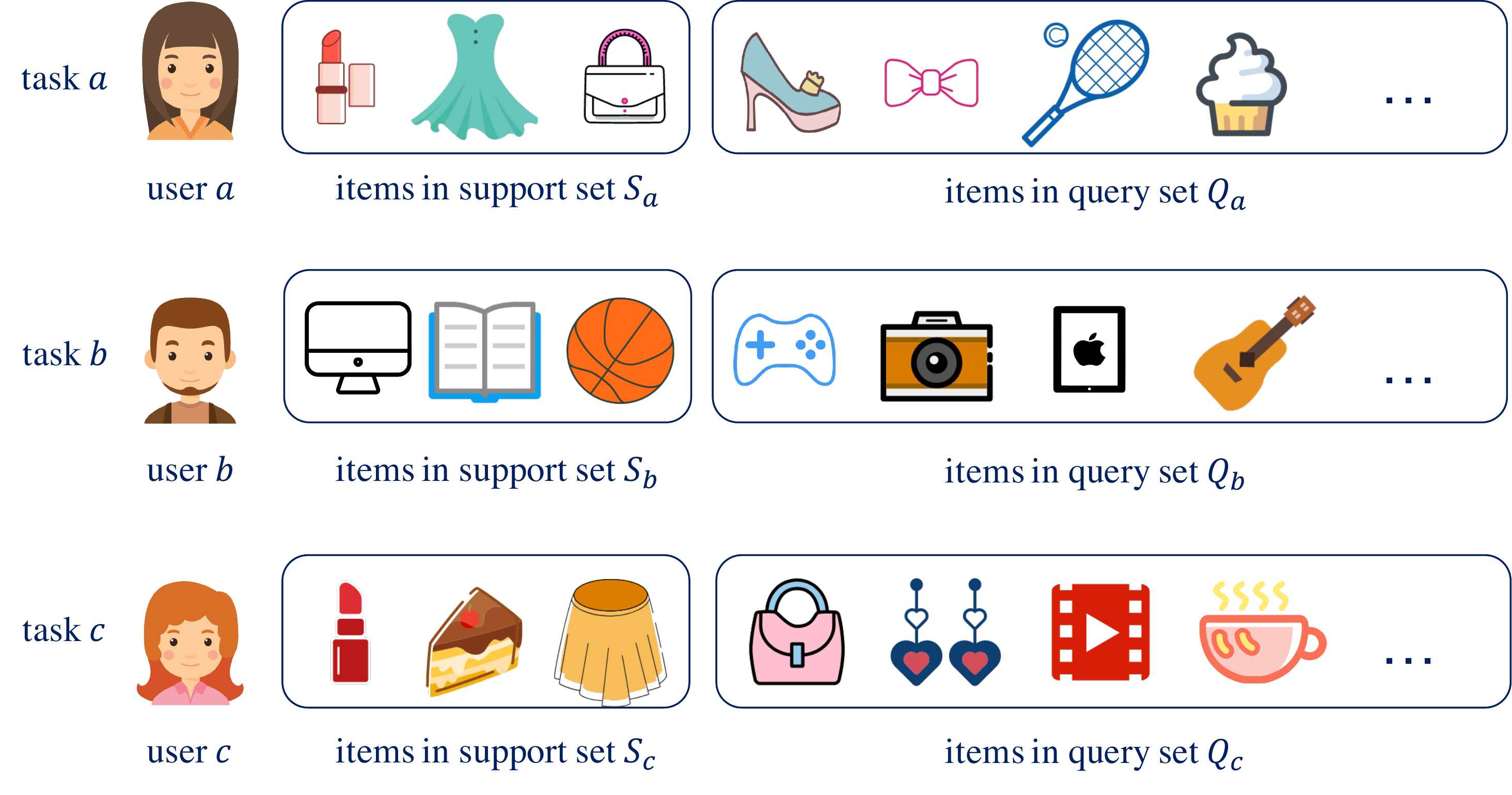}
\caption{An illustration of the relevance of different tasks. The purchase intentions of user $a$, $b$ and $c$ are manifested by the corresponding user-item interactions. It shows that tasks $a$, $c$ are closely relevant but task $b$ is largely different from them. Each task owns the user-specific support set and query set which will be explained in Definition \ref{def1}.}
\label{intro}
\end{center}
\vspace{-3mm}
\end{figure}

Most meta-learning recommenders are built upon the well-known framework of model-agnostic meta learning (MAML)~\cite{finn2017model}, aiming to learn a parameter initialization where a few steps of gradient updates will lead to good performances on the new tasks. A typical assumption here is the recommendations of different users are highly relevant. However, this assumption does not necessarily hold in actual scenarios. When the users exhibit different purchase intentions, the task relevance among them is actually very weak, which makes it problematic to find a shared parameter initialization optimal for all users. A concrete example is shown in Figure~\ref{intro}: tasks $a$ and $c$ share the transferable knowledge of recommendations, since user $a$ and user $c$ express similar purchase intentions, while task $b$ is largely different from them. Therefore, \emph{learning the relevance of different tasks} is an important step in adapting the global knowledge for the recommendations of cold-start users more effectively. 
\par

In this paper, we attempt to establish the connection between user cold-start recommendation and Neural Process (NP)~\cite{Garnelo2018NeuralP} to address above problem. 
NP is a neural-based approximation of stochastic processes. It is also related with meta learning~\cite{garnelo2018conditional}, since it can directly model the predictive distribution given a conditional prior learned from an \emph{arbitrary} number of context observations. To be specific, we propose a novel meta-learning framework called task-adaptive neural process (TaNP). Approximating each task as the particular instantiation of a stochastic process, our model is an effective ``few-shot function estimator'' to characterize the preference of cold-start user. TaNP performs amortized variational inference~\cite{kingma2013auto} to optimize the model parameters straightforwardly, which can relieve some inherent training issues in above MAML-based recommenders, such as model sensitivity~\cite{antoniou2018how} and being easily stuck into a local optimum~\cite{Dong2020MAMOMM}. 
\par
On top of that, TaNP includes a novel task-adaptive mechanism that is composed by a customized module and an adaptive decoder. In the customized module, the user interactions in each task are encoded as a deterministic task embedding which is interacted with the global pool to derive a clustering-friendly distribution. By resorting to the learned soft cluster assignments, we can capture the relevance of different tasks holistically. Afterwards, the customized module is combined with different modulation strategies to generate \emph{task-related} decoder parameters for making personalized recommendations. The main contributions of our work are summarized below: 
\begin{itemize}
    \item This paper proposes a formulation of tackling user cold-start recommendation within the neural process paradigm. Our model quickly learns the predictive distributions of new tasks, and the recommendations of cold-start users can be generated on the fly from the corresponding support set in the test phase.
    \item The novel introduction of task-adaptive mechanism does not only capture the relevance of different tasks but also incorporates the learned relevance into the modulation of decoder parameters, which is critical to better balance the trade-off between model capacity and adaptation reliability.
    \item Extensive experiments on benchmark datasets show that our model outperforms several state-of-the-art meta-learning recommenders consistently\footnote{The source code is available from \url{https://github.com/IIEdm/TaNP}.}. 
\end{itemize}

\section{Related Work}

\subsection{Meta Learning}
Meta learning covers a wide range of topics and has contributed to a booming study trend. Few-shot learning is one of successful branches of meta learning. We retrospect some representative meta-learning models with strong connections to our work. They can be divided into the following common types. 1) Memory-based approaches~\cite{santoro2016meta,kaiser2017learning}: combining deep neural networks (DNNs) with the memory mechanism to enhance the capability of storing and querying meta-knowledge. 2) Optimization-based approaches~\cite{ravi2016optimization,li2017meta}: a meta-learner, e.g. recurrent neural networks (RNNs) is trained to optimize target models. 3) Metric-based approaches~\cite{vinyals2016matching,snell2017prototypical}: learning an effective similarity metric between new examples and other examples in the training set. 4) Gradient-based approaches~\cite{finn2017model,nichol2018reptile}: learning an shared initialization where the model parameters can be trained via a few gradient updates on new tasks. Most meta-learning models follow an episodic learning manner. Among them, MAML is one of the most popular frameworks, which falls into the fourth type. Some MAML-based works~\cite{vuorio2019multimodal,yao2019hierarchically} consider that the sequence of tasks may originate from different task distributions, and try various task-specific adaptations to improve model capability.

\subsection{User Cold-Start Recommendation}
CF-based methods have been revolutionizing recommender systems in recent years due to the effectiveness of learning low-dimensional embeddings, like matrix factorization~\cite{gu2010collaborative,he2016fast} and deep learning~\cite{He2017NeuralCF,ma2019learning}. 
However, most of them are not explicitly tailored for solving user cold-start recommendation, e.g., new registered users only have very few interactions~\cite{park2009pairwise}. Previous methods~\cite{bianchi2017content,gao2018recommendation} mainly try to incorporate side-information into CF for alleviating this problem. 
\par
Inspired by the significant improvements of meta learning, the pioneering work~\cite{vartak2017meta} provides a meta-learning strategy to solve cold-start problems. It uses a task-dependent way to generate the varying biases of decision layers for different tasks, but it is prone to underfitting and is not flexible enough to handle various recommendation scenarios~\cite{bharadhwaj2019meta}. MeLU~\cite{lee2019melu} adopts the framework of MAML. Specifically, it divides the model parameters into two groups, i.e., the personalized parameter and the embedding parameter. The personalized parameter is characterized as a fully-connected DNN to estimate user preferences. The embedding parameter is referred as the embeddings of users and items learned from side-information. An inner-outer loop is used to update these two groups of parameters. In the inner loop, the personalized parameter is locally updated via the prediction loss of support set in current task. In the outer loop, these parameters are globally updated according to the prediction loss of query sets in multiple tasks. Through the fashion of local-global update, MeLU can provide a shared initialization for different tasks.
\par 
The later work MetaCS~\cite{bharadhwaj2019meta} is much similar to MeLU, and the main difference is that the local-global update  involves all parameters from input embedding to model prediction. 
To generalize well for different tasks, two most recent works MetaHIN~\cite{lu2020meta} and MAMO~\cite{Dong2020MAMOMM} propose different task-specific adaptation strategies. In particular, MetaHIN incorporates heterogeneous information networks (HINs) into MAML to capture rich semantics of meta-paths. MAMO introduces two memory matrices based on user profiles: a feature-specific memory that provides a specific bias term for the shared parameter initialization; a task-specific memory that guides the model for predictions. However, these two gradient-based meta-learning models may still suffer from potential training issues in MAML, and the model-level innovations of them are closely related with side-information, which limits their application scenarios.
\subsection{Neural Process}
NP is a neural-based formulation of stochastic processes that model the distribution over functions~\cite{Garnelo2018NeuralP,garnelo2018conditional}. It is a class of neural latent variable model that combines the strengths of Gaussian Process (GP) and DNNs to achieve flexible function approximations. 
NP mainly concentrates on the domains of low-dimensional function regression and uncertainty estimation~\cite{NIPS2019_9079}.
Meanwhile, there have been a growing number of researches on improving the expressiveness of vanilla NP model. For example, ANP~\cite{Kim2019AttentiveNP} introduces a self-attention NP to alleviate the underfitting of NP; CONVCNP~\cite{gordon2019convolutional} models the translation equivariance in data and extends task representations into function space. 
NP is also suitable to meta-learning problems~\cite{requeima2019fast}, because it provides an effective way to model the predictive distribution conditioned on a permutation-invariant representation learned from context observations. Our model is the first study that leverages the principle of NP to solve user cold-start recommendation, involving learn the discrete user-item relational data which is largely different from previous works. Moreover, TaNP includes a novel task-adaptive mechanism to better balance the trade-off between model capacity and adaptation reliability.
\begin{table}[!ht]
	\caption{Notations}
	\centering
	\label{tab:notation}
	\begin{tabular}{c|p{2in}}
		\hline
		Notation & Description\\
		\hline
		\hline
		$U$ and $V$ &user set and item set \\
	    $U^{tr}$ and $U^{te}$  &training user set and test user set \\
	    $\mathcal{T}^{tr}$ and $\mathcal{T}^{te}$  &training task set and test task set \\
	    $\tau_i$, $S_i$ and $Q_i$  &task $i$, support set $i$ and query set $i$ \\
	    $x_{i,j}=(u_i,v_j)$ &an interaction between $u_i$ and $v_j$ \\
	    $y_{i,j}$ & the actual rating of $x_{i,j}$\\ 
	    $N_i$ &the number of interactions in $\tau_i$ \\
	    $N_{S_{i}}$ and $N_{Q_{i}}$  &the numbers of interactions in $S_i$ and $Q_i$ \\
	    $\bm{c}_{i,n}$ & the one-hot content vector of $u_i$ \\
	    $\bm{E}_{n}$ & the shared matrix in embedding layers \\
	    $h_{\theta}$ & shared encoder for all tasks \\
	    $m_{\phi}$ & task identity network \\
	    $\bm{A}$ & global pool \\ 
	    $\bm{C}$ & soft cluster assignment matrix \\ 
	    $\bm{D}$ & clustering target distribution \\
	    $g_{\omega_{i}}$  & adaptive decoder for $\tau_i$ \\
	    $\bm{\gamma}^{l}_{i}$, $\bm{\beta}^{l}_{i}$, $\bm{\eta}^{l}_{i}$ and $\bm{\delta}^{l}_{i}$ & feature modulation parameters of $g_{\omega_{i}}$ for the $l$-th layer \\
		\hline
		\end{tabular}
\end{table}
\section{Preliminaries}
We formulate the problem of user cold-start recommendation from the meta-learning perspective. Following the traditional episodic learning manner, we first give the formal definition of task (episode).
\begin{definition}\label{def1}
\textbf{(TASK)}
Given the user set $U$, the item set $V$ and a specific user $u_{i} \in U$, making the personalized recommendation for $u_{i}$ is defined as a task $\tau_{i}$. $\tau_{i}$ includes the available interactions of $u_{i}$, i.e., $\{x_{i,j}, y_{i,j}\}_{j=1}^{N_i}$. $x_{i,j}$ denotes a tuple $(u_{i}, v_{j})$ in which $v_{j} \in V$ denotes an item and $y_{i, j}$ is $u_{i}$'s actual rating of item $v_{j}$. $N_{i}$ denotes the number of interactions in $\tau_{i}$. For notation simplicity, we also denote $\tau_{i} = \{x_{i,j}, y_{i,j}\}_{j=1}^{N_i}$. $\tau_{i}$ contains few interactions as the support set $S_{i}$ and remaining interactions as query set $Q_{i}$, i.e., $\{x_{i,j}, y_{i,j}\}_{j=1}^{N_i} = S_{i} \cup Q_{i}$. Here we denote $S_{i} = \{x_{i,j}, y_{i,j}\}^{N_{S_{i}}}_{j=1}$ and $Q_{i} = \{x_{i,j}, y_{i,j}\}^{N_{S_{i}}+N_{Q_{i}}}_{j=N_{S_{i}}+1}$. $N_{S_{i}}$ denotes the number of interactions in $S_{i}$, which is typically set with a small value, and $N_{Q_{i}}$ is the number of interactions in $Q_{i}$ with $ N_{S_{i}} +N_{Q_{i}}= N_{i}$.
\end{definition}
Note a meta-learning recommender is first learned on each support set (learning procedure) and is then updated according to the prediction loss over multiple query sets (learning-to-learn procedure). Through the guide of the second procedure in many iterations, 
this meta-learning model can derive the global knowledge across different tasks and adapts such knowledge well for a new task $\tau_i$ with only $S_i$ available. 
\par
For this purpose, we split $U$ into two disjoint sets: training user set $U^{tr}$ and test (cold-start) user set $U^{te}$. The set of all training tasks is denoted as $\mathcal{T}^{tr} =\{\tau_{i} | u_i \in U^{tr}\}$ and the set of all test tasks is denoted as $\mathcal{T}^{te} = \{\tau_{i} | u_i \in U^{te} \}$. 
In the training phase, we can train our model following a learning-to-learn manner for each training task. In the test phase, when a cold-start user $u_{i'} \in U^{te}$ comes, our goal is to find which items $u_{i'}$ will be interested in, based on a small number of interactions in $S_{i'}$ and the global knowledge learned from $\mathcal{T}^{tr}$. 
The notations used in this paper are summarized in Table 1. In addition, according to the concrete value of $y_{i,j}$, the personalized recommendation can be either viewed as a binary classification to indicate whether $u_{i}$ engages with $v_{j}$, or a regression problem that $v_{j}$ has different ratings that need to be assessed by $u_{i}$. 
\section{Task-adaptive Neural Process}
In this section, we first describe how to handle user cold recommendation from the view of NP. Then we analysis the potential issue of it and propose an effective solution, i.e., the task-adaptive mechanism. Our model includes three parts: encoder, customization module and adaptive decoder. The encoder is used to obtain the variational prior and posterior via a lower bound estimation. The customization module is used to recognize the task identity and to learn the relevance of different tasks. It is further coupled with different modulation strategies to generate the model parameters of adaptive decoder. In addition, our model includes different embedding strategies and prediction losses, which can be applicable to different recommendation scenarios.
\subsection{Overview}
In our model, we assume each task $\tau_{i} = \{x_{i,j}, y_{i,j}\}_{j=1}^{N_i}$ is associated with an instantiation of stochastic process $f_{i}$ from which the observed interactions of user $u_i$ are drawn. The corresponding joint distribution $\rho$ can be given as:
\begin{equation}
\begin{split}
\rho_{x_{i, 1:N_{i}}}(y_{i, 1:N_{i}}) = \int p(f_{i})p(y_{i, 1:N_{i}} | f_i, x_{i, 1:N_{i}})df_i 
\label{joint_distribution}
\end{split}
\end{equation}
Here we use $x_{i, 1:N_{i}}$ and $y_{i, 1:N_{i}}$ to denote $\{x_{i,j}\}_{j=1}^{N_{i}}$ and $\{y_{i,j}\}_{j=1}^{N_{i}}$ decoupled from $\tau_{i}$. Motivated by NP, we can approximate the above stochastic process via a fixed-dimensional vector $\bm{z}_{i}$ and the learnable non-linear functions parameterized by DNNs. 
The complete generation process with the i.i.d. condition can be given as follows, 
\begin{equation}
\begin{split}
p(y_{i, 1: N_{i}}|x_{i, 1: N_{i}}) = \int p(\bm{z}_{i}) \prod_{j=1}^{N_{i}} p(y_{i,j}|x_{i,j}, \bm{z}_{i}) d\bm{z}_{i}.
\label{generation_process}
\end{split}
\end{equation}
In such a way, sampling $\bm{z}_{i}$ from $p(\bm{z}_{i})$ can be viewed as a concrete function realization. 
\par
Since the true posterior is intractable, we use amortized variational inference~\cite{mnih2014neural} to learn it. The variational posterior of the latent variable $\bm{z}_{i}$ is defined as $q(\bm{z}_{i}|\tau_{i})$, and we have the following evidence lower-bound (ELBO) objective with step-by-step derivations:
\begin{equation}
\begin{split}
&{\rm log} p(y_{i, 1: N_{i}}|x_{i, 1: N_{i}}) = {\rm log}\int p(\bm{z}_{i}, y_{i, 1: N_{i}}|x_{i, 1: N_{i}}) d\bm{z}_{i} \\
&={\rm log}\int p(\bm{z}_{i}, y_{i, 1: N_{i}}|x_{i, 1: N_{i}}) \frac{q(\bm{z}_{i}|\tau_{i})}{q(\bm{z}_{i}|\tau_{i})} d\bm{z}_{i} \\
&\geq \mathbb{E}_{q(\bm{z}_{i}|\tau_{i})} \big[{\rm log}\frac{p(\bm{z}_{i}, y_{i, 1: N_{i}}|x_{i, 1: N_{i}})}{q(\bm{z}_{i}|\tau_{i})} \big] \\
&= \mathbb{E}_{q(\bm{z}_{i}|\tau_{i})} \big [{\rm log}\frac{p(\bm{z}_{i}) p(y_{i, 1: N_{i}}|x_{i, 1: N_{i}}, \bm{z}_{i})}{q(\bm{z}_{i}|\tau_{i})} \big] \\
&=\mathbb{E}_{q(\bm{z}_{i}|\tau_{i})}\big[\sum_{j = 1}^{N_{i}}{\rm log}p(y_{i,j}|x_{i, j}, \bm{z}_{i}) + {\rm log}\frac{p(\bm{z}_{i})}{q(\bm{z}_{i}|\tau_{i})} \big].
\end{split}
\label{elbo}
\end{equation}
Each $\tau_{i}$ is constituted by $S_{i}$ and $Q_{i}$, and we should pay more attention to make predictions for $Q_{i}$ given $S_{i}$. Therefore, Eq.(\ref{elbo}) is alternatively defined as: 
\begin{equation}
\begin{split}
&{\rm log} p(y_{i, 1: N_{Q_{i}}}|x_{i, 1: N_{Q_{i}}}, S_{i}) \\
&\geq \mathbb{E}_{q(\bm{z}_{i}|\tau_{i})}\big[\sum_{j = 1}^{N_{Q_{i}}}{\rm log}p(y_{i, j}|x_{i, j}, \bm{z}_{i}) + {\rm log}\frac{q(\bm{z}_{i}|S_{i})}{q(\bm{z}_{i}|\tau_{i})} \big].
\end{split}
\label{elbo2}
\end{equation}
We use the variational prior $q(\bm{z}_{i}|S_{i})$ to approximate $p(\bm{z}_{i}|S_{i})$ in Eq.(\ref{elbo2}), since $p(\bm{z}_{i}|S_{i})$ is also intractable. Such an approximation can be considered as a regularization term of KL divergence to approximate the \emph{consistency} condition of stochastic process~\cite{oksendal2013stochastic}. Furthermore, to model the distribution over random functions, we add the variability of interaction sequence to $\tau_{i}$ as suggested in~\cite{Garnelo2018NeuralP}. Then, each sample of $\bm{z}_{i}$ is regarded as a realisation of the corresponding stochastic process.
\par
\begin{figure*}
\begin{center}
\includegraphics[width=17cm,height=8cm]{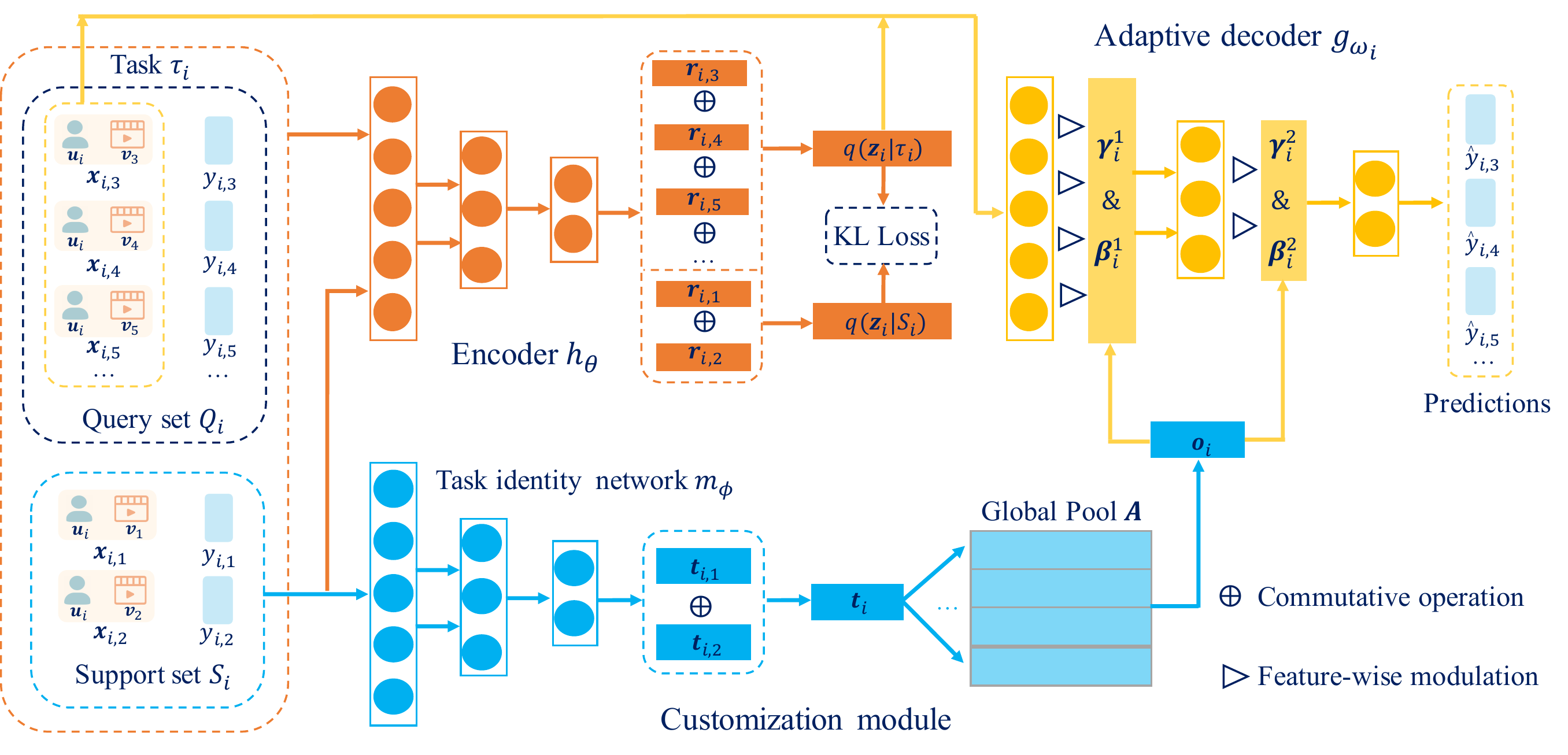}
\caption{The framework of TaNP in the training phase. TaNP includes the encoder $h_{\theta}$, the customization module (task identity network $m_{\phi}$ and global pool $\bm{A}$) and the adaptive decoder $g_{\omega_{i}}$. Both of $S_i$ and $\tau_i$ are encoded by $h_\theta$ to generate the variational prior and posterior, respectively. The final task embedding $\bm{o}_{i}$ learned from the customized module is used to modulate the model parameters of $g_{\omega_{i}}$. $\bm{z}_i$ sampled from $q(\bm{z}_{i}|\tau_{i})$ is concatenated with $\bm{x}_{i, j}$ to predict $\hat{y}_{i,j}$ via $g_{\omega_{i}}$.}
\label{model}
\end{center}
\end{figure*}

From above formulations, our model can be extended to learn multiple tasks with different stochastic processes in a meta-learning framework. However, both the \emph{encoder} $q(\bm{z}|\tau)$/$q(\bm{z}|S)$ and the \emph{decoder} $p(y|\bm{z},x)$ are the global parameters that are shared by all training tasks $\mathcal{T}^{tr}$, and the relevance of different tasks has been ignored. Hence, the above framework is inflexible to adjust model capacity for different tasks. In other words, it may lead to underfitting for some relative complex tasks, and suffers from overfitting for some easy tasks. To better balance the trade-off between model capacity and adaptation reliability, we introduce a novel task-adaptive mechanism. The overall training framework of our model is shown in Figure~\ref{model}.
\subsection{Embedding Layers}
We first use the embedding layers to generate initial user and item embeddings as the inputs of TaNP. Our model is compatible with side-information of users and items, thus we provide two embedding strategies. Taking a user $u_i$ for example, when categorical contents of $u_i$ are available, we would generate a content embedding for each categorical content and concatenate them together to obtain the initial user embedding. Given $n$ user contents, the embedding procedure is defined as: 
\begin{equation}
\begin{split}
\bm{u}_{i} = \big[\bm{E}_{1}\bm{c}_{i,1} | ... | \bm{E}_{n}\bm{c}_{i,n}\big],
\label{embed_1}
\end{split}
\end{equation}
where $[\cdot|\cdot]$ is the concatenation operation, $\bm{c}_{i,n}$ is the one-hot vector of $n$-th categorical content of $u_{i}$, and $\bm{E}_{n}$ represents the corresponding shared embedding matrix. When such kind of user contents are not available, the user embedding can be obtained by
\begin{equation}
\begin{split}
\bm{u}_{i} = \sigma(\bm{W}_{2}\sigma(\bm{W}_{1}\bm{e}_{i}+\bm{b}_1) + \bm{b}_2),
\label{embed_2}
\end{split}
\end{equation}
where $\{\bm{W}_{1}, \bm{W}_{2}\}$ denotes weight matrices, $\{\bm{b}_{1}, \bm{b}_2\}$ denotes bias vectors, $\sigma$ is the sigmoid activation function, and $\bm{e}_{i}$ is the one-hot vector of $u_{i}$. Notice that the initial item embeddings are obtained by following the similar embedding process.
\subsection{Encoder}
\par
Given $S_{i}$ and $\tau_{i}$, our encoder tries to generate the variational approximations $q(\bm{z}_{i}|S_{i})$ and $q(\bm{z}_{i}|\tau_{i})$ respectively. Concretely, for an interaction $(x_{i,j}, y_{i,j})$ in  $S_{i}$ or $\tau_{i}$ , the encoder $h_{\theta}$ would produce a corresponding embedding $\bm{r}_{i,j}$ via the following operation:
\begin{equation}
\begin{split}
\bm{r}_{i,j} = h^{l}(h^{l-1}(\cdot \cdot \cdot h^{1}([\bm{u}_{i} | \bm{v}_{j} | y_{i,j}]))),
\label{embed_r}
\end{split}
\end{equation}
where $h(\bm{x}) = {\rm ReLU}(\bm{Wx}+\bm{b})$ is a fully-connected layer with the corresponding $\bm{W}$ and $\bm{b}$, and $l$ is the number of stacked layers. Given a set of observed interactions, the encoder would then aggregate these encoded vectors to generate a permutation-invariant representation $\bm{r}_{i}$. For example, to model $q(\bm{z}_{i}|\tau_{i})$, we have
\begin{equation}
\begin{split}
\bm{r}_{i} = \bm{r}_{i,1} \oplus \bm{r}_{i,2} \oplus \ ...\ \bm{r}_{i,N_{i-1}} \oplus \bm{r}_{i,N_{i}},
\label{aggregation}
\end{split}
\end{equation}
where $\oplus$ is a commutative operation and we use a mean operation, i.e., $\bm{r}_{i} = \frac{1}{N_{i}} \sum_{j=1}^{N_{i}} \bm{r}_{i,j}$ for efficiency. This permutation invariance in Eq.(\ref{aggregation}) is also an important step to approximate the \emph{exchangeability} condition of stochastic process~\cite{oksendal2013stochastic}. The reparameterization trick~\cite{kingma2013auto} is adopted to express the random variable, i.e., $\bm{z}_{i} \thicksim \mathcal{N}(\bm{\mu}_{i},{\rm diag} (\bm{\sigma}_{i}^{2}))$. It can be formally defined as:
\begin{equation}
\begin{split}
\bm{r}_{i} = ~&{\rm ReLU}(\bm{W}_{s} \bm{r}_{i}), \\ 
\bm{\mu}_{i} = \bm{W}_{\mu} &\bm{r}_{i}, \
{\rm log}\bm{\sigma}_{i} = \bm{W}_{\sigma} \bm{r}_{i}, \\
\bm{z}_{i} = \bm{\mu}_{i} + &\bm{\epsilon} \odot \bm{\sigma}_{i},\ \bm{\epsilon} \thicksim \mathcal{N}(\bm{0}, \bm{I}),
\label{re_trick}
\end{split}
\end{equation}
where $\odot$ denotes the element-wise product, $\bm{\epsilon}$ is the Gaussian noise, and $\{\bm{W}_{s}, \bm{W}_{\mu}, \bm{W}_{\sigma} \}$ are weight matrices. 
\subsection{Customization Module}
The customization module seeks to learn the relevance of different tasks. It contains a task identity network $m_{\phi}$ and a global pool $\bm{A}$. $m_{\phi}$ is used to produce a temporary task embedding $\bm{t}_{i}$ from the corresponding support set $S_{i}$. 
Formally, it encodes each interaction $(x_{i,j}, y_{i,j})$ in $S_{i}$ as a low-dimensional representation $\bm{t}_{i,j}$, which can be described as:
\begin{equation}
\begin{split}
\bm{t}_{i,j} = m^{l}(m^{l-1}(\cdot \cdot \cdot m^{1}([\bm{u}_{i} | \bm{v}_{j} | y_{i,j}]))).
\label{embed_t}
\end{split}
\end{equation}
$m_{\phi}$ keeps the same network structure with Eq.(\ref{embed_r}). $\{\bm{t}_{i,j}\}_{j=1}^{N_{S_{i}}}$ is aggregated into $\bm{t}_{i}$ via the same operation in Eq.(\ref{aggregation}). The global pool $\bm{A} = [\bm{a}_{1}, ..., \bm{a}_{k}] \in R^{d \times k}$ is a differentiable external resource that preserves the soft c ter centroids. Each $\bm{t}_{i}$ would interact with $\bm{A}$ to derive soft cluster assignments ($k$ is a hyper-parameter that represents the number of soft cluster centroids). We use the Student's t-distribution as a kernel to measure the normalized similarity between $\bm{t}_{i}$ and $\bm{a}_{j}$ as follows,
\begin{equation}
\begin{split}
c_{i,j} = \frac{(1+||\bm{t}_{i} - \bm{a}_{j}||^{2}/\alpha)^{-\frac{\alpha+1}{2}}}{\sum_{j'}(1+||\bm{t}_{i} - \bm{a}_{j'}||^{2}/\alpha)^{-\frac{\alpha+1}{2}}},
\label{cluster}
\end{split}
\end{equation}
where $\alpha$ is the degree of freedom of the Student's t-distribution. The final task embedding $\bm{o}_{i}$ is generated by the following operation:
\begin{equation}
\begin{split}
\bm{o}_{i} = \sigma(\bm{W}_{o}(\bm{t}_{i} + \bm{A}\bm{c}_{i}^{T})),
\label{final_task}
\end{split}
\end{equation}
where $\bm{W}_{o}$ is a weight matrix. In fact, the sequence of interactions in each task represents the purchase intentions of a specific user, and different users may have similar or diverse intentions. $\bm{A}$ includes multiple high-level features that are related with user intentions, which is similar to the intention prototypical embeddings used in disentangled recommendation models~\cite{ma2019learning,ma2020disentangled}. Each task has access to $\bm{A}$, and the correlation degree is reflected by $\bm{c}_{i}$. Therefore, the relevance of different tasks can be globally learned, which is further incorporated into the final task embedding through the Eq.(\ref{final_task}). 
\par
The normalized assignments of all training tasks construct a assignment matrix $\bm{C} = [\bm{c}_{1}, ..., \bm{c}_{|\mathcal{T}^{tr}|}] \in R^{|\mathcal{T}^{tr}| \times k}$. As suggested by~\cite{xie2016unsupervised}, we use an unsupervised clustering loss $\mathcal{L}_{u}$ with the guidance of an auxiliary clustering target distribution $\bm{D}$. $\mathcal{L}_{u}$ is defined as a KL divergence loss between $\bm{C}$ and $\bm{D}$:
\begin{equation}
\begin{split}
\mathcal{L}_{u} = {\rm KL}(\bm{D}||\bm{C}) = \sum_{i}\sum_{j}\bm{D}_{i,j}{\rm log}\frac{\bm{D}_{i,j}}{\bm{C}_{i,j}}. 
\label{cluster_loss}
\end{split}
\end{equation}
where the clustering target distribution $\bm{D}$ can be defined as follows,
\begin{equation}
\begin{split}
\bm{D}_{i,j} = \frac{(\bm{C}_{i,j})^2 / \sum_{i}\bm{C}_{i,j}}{\sum_{j'}(\bm{C}_{i,j'})^{2} / \sum_{i}\bm{C}_{i,j'}}. 
\label{target_distribution}
\end{split}
\end{equation}
Our clustering improves cluster purity and puts more emphasis on tasks assigned with high confidence.
\subsection{Adaptive Decoder}
The original decoder $g_{\omega}$ in Eq.(\ref{elbo2}) is used to learn the conditional likelihood $p(y|x,z)$, which is a global predictor shared by all tasks. In this section, TaNP introduces an adaptive decoder $g_{\omega_{i}}$ in a parameter-efficient manner. We describe two candidate modulation strategies to generate the model parameters of adaptive decoder via using the final task embedding $\bm{o}_{i}$. The first variant is Feature-wise Linear Modulation (FiLM)~\cite{perez2018film}. Based on this basic modulation, we propose the second modulation strategy, i.e., Gating-FiLM.   
\subsubsection{FiLM} FiLM tires to adaptively influence the prediction of a DNN by applying a feature-wise affine transformation on its intermediate features. It has been proved to be highly effective in many domains~\cite{cadene2019murel,brockschmidt2019gnn}. Here we employ FiLM to scale and shift the feature of each layer of our decoder via two generated parameters. The adaptation of $g_{\omega_{i}}$ for the $l$-th layer can be defined as:
\begin{equation}
\begin{split}
\bm{\gamma}^{l}_{i} = {\rm tanh}(\bm{W}_{\gamma}^{l}\bm{o}_{i}), \quad &\bm{\beta}^{l}_{i} = {\rm tanh}(\bm{W}_{\beta}^{l}\bm{o}_{i}), \\
\bm{g}_{i,j}^{l+1} = {\rm ReLU}(\bm{\gamma}^{l}_{i} \odot (&\bm{W}_{\omega}^{l} \bm{g}_{i,j}^{l} + \bm{b}_{\omega}^{l}) + \bm{\beta}^{l}_{i}), 
\label{adaption}
\end{split}
\end{equation}
where $\{\bm{W}_{\gamma}^{l}, \bm{W}_{\beta}^{l}, \bm{W}_{\omega}^{l}\}$ denotes layer-wise weight matrices, $\bm{b}_{\omega}^{l}$ denotes a bias vector, and $\bm{g}_{i,j}^{l}$ is the input of $l$-th layer of our decoder. The non-linearity function $\rm tanh$ is applied here to restrict the output of modulation to be in $[-1, 1]$. In the first layer, the input of $g_{\omega_{i}}$ is the concatenated vector of $\bm{x}_{i, j}$ and $\bm{z}_{i}$, i.e., $\bm{g}^{1}_{i,j} = [\bm{u}_{i}|\bm{v}_{j}|\bm{z}_{i}]$.

\subsubsection{Gating-FiLM} Although FiLM is effective to achieve the feature modulation, a potential weakness is that such operation cannot filter some information which has the opposite effects on learning. To alleviate this problem, we introduce a gating version of FiLM:
\begin{equation}
\begin{split}
\bm{\gamma}^{l}_{i} &= {\rm tanh}(\bm{W}_{\gamma}^{l}\bm{o}_{i}),\  \bm{\beta}^{l}_{i} = {\rm tanh}(\bm{W}_{\beta}^{l}\bm{o}_{i}),\\
\bm{\eta}^{l}_{i} &= {\rm tanh}(\bm{W}_{\eta}^{l}\bm{o}_{i}),\ \bm{\delta}^{l}_{i} = \sigma(\bm{W}_{\delta}^{l}\bm{o}_{i}), \\
&\bm{\gamma}^{l}_{i} = \bm{\gamma}^{l}_{i} \odot \bm{\delta}^{l}_{i}
+ \bm{\eta}^{l}_{i} \odot (\bm{1} - \bm{\delta}^{l}_{i}), \\
&\bm{\beta}^{l}_{i} = \bm{\beta}^{l}_{i} \odot \bm{\delta}^{l}_{i} + \bm{\eta}^{l}_{i} \odot (\bm{1} - \bm{\delta}^{l}_{i}) , \\
\bm{g}_{i,j}^{l+1} &= {\rm ReLU}(\bm{\gamma}^{l}_{i} \odot (\bm{W}_{\omega}^{l} \bm{g}_{i,j}^{l} + \bm{b}_{\omega}^{l}) + \bm{\beta}^{l}_{i}), 
\label{adaption2}
\end{split}
\end{equation}

\begin{algorithm}
\caption{The training procedure of TaNP.}
\label{alg}
\begin{algorithmic}[1]
\REQUIRE Training user set $U^{tr}$; Item set $V$; User and item side-information (optional); Training task set $\mathcal{T}^{tr}$; Hyper-parameters: $d$, $l$, $k$, $\alpha$, $\lambda$.
\ENSURE Parameters in embedding layer; $h_{\theta}$; $m_{\phi}$; $\bm{A}$; $g_{\omega_{i}}$.
\STATE Initialize all model parameters.
\WHILE {not convergence} 
\FOR {$\tau_{i} \in \mathcal{T}^{tr}$}
\STATE Construct $S_i$ and $Q_i$ from $\tau_{i}$.
\STATE Generate $q(\bm{z}_{i}|\tau_{i})$ via $h_{\theta}$ in Eq.(\ref{re_trick}).
\STATE Generate task embedding $\bm{o}_{i}$ via $m_{\phi}$ and $\bm{A}$ in Eq.(\ref{embed_t})-(\ref{final_task}).
\STATE Predictions on $Q_{i}$ via adaptive decoder $g_{\omega_{i}}$, $\bm{z}_{i}$ and $\bm{o}_{i}$ in Eq.(\ref{adaption}) or Eq.(\ref{adaption2}).
\STATE Generate $q(\bm{z}_{i}|S_{i})$ via $h_{\theta}$ in Eq.(\ref{re_trick}).
\STATE Calculate prediction loss $\mathcal{L}_{r,i}$ in  Eq.(\ref{regression_loss}) or Eq.(\ref{binary_loss}).
\STATE Calculate regularization loss $\mathcal{L}_{c,i}$ in Eq.(\ref{total_loss}).
\ENDFOR
\STATE Calculate clustering loss $\mathcal{L}_u$ in Eq.(\ref{cluster_loss}) and the total loss $\mathcal{L}$ in Eq.(\ref{total_loss}).
\STATE Update model parameters by Adam optimizer.
\ENDWHILE
\end{algorithmic}
\end{algorithm}
where \{$\bm{W}_{\eta}^{l}$, $\bm{W}_{\delta}^{l}$\} are two weight matrices and $\bm{\delta}^{l}_{i}$ is the gating term to control the influences of $\bm{\gamma}^{l}_{i}$ and $\bm{\beta}^{l}_{i}$. FiLM and Gating-FiLM can be alternatively used in our adaptive decoder, and the experiments verify their effectiveness.

\subsection{Loss Function}
In our model, the likelihood term in Eq.(\ref{elbo2}) is reformulated as a regression-based loss function: 
\begin{equation}
\begin{split}
\mathcal{L}_{r,i} &= -\mathbb{E}_{q(z_{i}|\tau_{i})}{\rm log} p(y_{i, 1: N_{Q_{i}}}|x_{i, 1: N_{Q_{i}}}, \bm{z}_{i}) \\
& \propto \frac{1}{N_{Q_{i}}} \sum_{j =1}^{N_{Q_{i}}}(y_{i,j} - \hat{y}_{i,j})^{2}, 
\label{regression_loss}
\end{split}
\end{equation}
where $\hat{y}_{i,j}$ is the final output of $g_{\omega_{i}}(x_{i,j},\bm{z}_{i}, \bm{o}_{i})$. 
For implicit feedback data, $\mathcal{L}_{r,i}$ is defined as a binary cross-entropy loss:
\begin{equation}
\begin{split}
\mathcal{L}_{r,i} &= -\mathbb{E}_{q(z_{i}|\tau_{i})}{\rm log} p(y_{i, 1: N_{Q_{i}}}|x_{i, 1: N_{Q_{i}}}, \bm{z}_{i}) \\
\propto -\frac{1}{N_{Q_{i}}} & \sum_{j =1}^{N_{Q_{i}}} y_{i,j}{\rm log}(\hat{y}_{i,j}) + (1-y_{i,j}){\rm log}(1-\hat{y}_{i,j}). 
\label{binary_loss}
\end{split}
\end{equation}
The training loss of TaNP is defined as:
\begin{equation}
\begin{split}
\mathcal{L} = \frac{1}{|\mathcal{T}^{tr}|}\sum_{i=1}^{|\mathcal{T}^{tr}|}(\mathcal{L}_{r,i} + \mathcal{L}_{c,i}) + \lambda \mathcal{L}_{u},
\label{total_loss}
\end{split}
\end{equation}
where $\mathcal{L}_{c,i} = {\rm KL} (q(z_{i}|\tau_{i})||q(z_{i}|S_{i}))$ that can be also considered as a regularization term to approximate the condition of consistency, and $\lambda$ is a hyper-parameter which is selected between 0 and 1. Our model is an end-to-end framework which can be optimized by Adam~\cite{Kingma2015AdamAM} empirically. 
The pseudo code of training procedure is given in Algorithm~\ref{alg}.
\par
It should be noticed that the test procedure of our model is different from Algorithm~\ref{alg}. Concretely, in the test phase, the ground truth $y_{i,j}$ in $Q_i$ is not available, thus $\bm{z}_{i}$ is sampled from $q(\bm{z}_{i} | S_{i})$ via our encoder $h_{\theta}$. By the same token, our final task embedding $\bm{o}_{i}$ is always obtained from the available interactions in $S_i$ instead of $\tau_i$. After that, $\bm{o}_{i}$ is used to modulate the model parameters of $g_{\omega_{i}}$, and $\bm{z}_{i}$ is concatenated with $\bm{x}_{i,j}$ to predict $\hat{y}_{i,j}$ for $Q_i$.
\subsection{Time Complexity}
In TaNP, $h_{\theta}$, $m_{\phi}$ and $g_{\omega_{i}}$ are parameterized as fully-connected DNNs. Therefore, our model keeps an efficient architecture for training and inference. The time complexity of each component can be approximated as $O(ld^3)$. Here we use $d$ to represent the hidden size of each layer. In fact, $d$ is varied among different layers. The calculation of final task embedding $\bm{o}_i$ would cost $O(kd^2)$. As FiLM/Gating-FiLM only requires two/four weight matrices per layer in $g_{\omega_{i}}$, both of them are computationally efficient adaptations. Furthermore, our model can be also optimized in a batch-wise manner for parallel computations. 
\begin{table}
\centering
\caption{Statistics of datasets.}
\resizebox{8.5cm}{!}{
\begin{tabular}{|l|c|c|c|c|c|}
\hline
\textbf{Datasets}  & \textbf{Users} & \textbf{Items} & \textbf{Ratings} &\textbf{Type} &\textbf{Content}\\ \hline
MovieLens-1M  &6,040  & 3,706 &1,000,209  &explicit &yes\\ \hline
Last.FM &1,872 &3,846 &42,346 &implicit &no\\ \hline
Gowalla &2692  &27,237  &134,476  &implicit &no\\ \hline
\end{tabular}
}
\label{dataset}
\end{table}
\section{Experiments}
In this section, we seek to answer the following major research questions (RQs):
\begin{itemize}
    \item \textbf{RQ1}: Does our method achieve the supreme performances in comparison with other cold-start models, including 1) the classic cold-start models and CF-based DNN models, 2) the popular meta-learning models, 3) an ablation of our method where the proposed task-adaptive mechanism is removed and 4) different modulation strategies, i.e., FiLM and Gating-FiLM used in $g_{\omega_{i}}$ (Section 5.3)?
    \item \textbf{RQ2}: When we change the number of interactions in support set for each task, i.e., $N_{S_i}$, the available information in the test phase is reduced. Is our model still able to achieve fast adaptations for cold-start users (Section 5.4)? 
    \item \textbf{RQ3}: What is learned by our customization module? Is the relevance of different tasks well captured (Section 5.5)? 
    \item \textbf{RQ4}: How well does the proposed method generalize when we tune different hyper-parameters for it (Section 5.6)? 
\end{itemize}
To answer these questions, we do a detailed comparative analysis of our model on public benchmark datasets.
\subsection{Datasets}
We evaluate TaNP on three real-world recommendation datasets: MovieLens-1M\footnote{https://grouplens.org/datasets/movielens/1m/}, Last.FM\footnote{https://grouplens.org/datasets/hetrec- 2011/} and Gowalla\footnote{http://snap.stanford.edu/data/loc-gowalla.html}. MovieLens-1M is a widely used movie dataset with explicit ratings (from 1 to 5). Last.FM is a music dataset that contains musician listening information from users in Last.fm online music system, and we directly use it provided by~\cite{wang2019multi}. Gowalla is a location-based social network that contains user-venue checkins, and we extract a part of interactions from it. The details of these datasets are summarized in Table~\ref{dataset}. 
\begin{table*}[]
\centering
\caption{Performance (\%) comparison of user cold-start recommendation on MovieLens-1M.}
\resizebox{17.5cm}{!}{
\begin{tabular}{cccccccccc}
\hline
Model &P@5 &NDCG@5 &MAP@5 &P@7 &NDCG@7 &MAP@7 &P@10 &NDCG@10 &MAP@10 \\ \hline \hline
PPR &49.36 &66.62 &37.86 &51.25 &67.27 &38.12 &54.30 &68.27 &41.20 \\ 
NeuMF &48.27 &65.43 &36.65 &51.24 &66.55 &37.90 &55.32 &68.19 &41.43 \\
DropoutNet &51.77 &69.34 &41.82 &53.67 &70.83 &43.81 &57.34 &72.02 &46.59 \\
MeLU &55.99 &73.08 &46.79 &57.34 &73.18 &48.45 &61.05 &74.04 &49.02 \\ 
MetaCS &55.43 &71.69 &44.85 &56.89 &72.05 &44.94 &59.78 &72.86 &47.52 \\ 
MetaHIN  &57.65 &73.43 &47.40 &58.67 &73.95 &48.75 &61.18 &74.50 &49.99 \\
MAMO  &57.69 &73.24 &47.72 &58.42 &74.03 &49.62 &61.51 &74.41 &50.06 \\ \hline
TaNP (w/o tm) &58.03 &73.76 &47.79 &58.90 &73.89 &48.37 &61.29 &74.44 &49.73 \\ 
TaNP (FiLM) &59.76 &74.97 &49.08 &$\bm{60.45}$ &75.22 &49.76 &$\bm{62.78}$ &75.48 &51.12 \\
TaNP (Gating-FiLM) &$\bm{60.12}$ &$\bm{75.00}$ &$\bm{49.12}$  &60.29 &$\bm{75.34}$ &$\bm{50.79}$ &62.66 &$\bm{75.53}$ &$\bm{51.56}$  \\ 
\hline
\end{tabular}
}
\label{movie}
\end{table*}

\begin{table*}[]
\centering
\caption{Performance (\%) comparison of user cold-start recommendation on Last.FM.}
\resizebox{17.5cm}{!}{
\begin{tabular}{cccccccccc}
\hline
Model &P@5 &NDCG@5 &MAP@5 &P@7 &NDCG@7 &MAP@7 &P@10 &NDCG@10 &MAP@10 \\ \hline \hline
PPR &68.61 &67.23 &61.72 &72.96 &69.10 &67.29 &81.36 &72.75 &75.66 \\
NeuMF &67.39 &65.23 &59.72 &70.82 &67.62 &67.80 &81.06 &71.57 &76.01 \\
DropoutNet &70.08 &68.37 &62.68 &73.34 &69.72 &68.66 &82.28 &73.26 &79.84 \\
MetaLWA &69.45 &69.04 &62.31 &73.38 &70.92 &69.29 &86.06 &74.78 &82.58 \\
MetaNLBA &71.52 &72.47 &64.71 &75.97 &73.15 &71.06 &86.52 &78.40 &83.51 \\
MeLU &73.03 &75.38 &67.71 &76.19 &75.54 &72.09 &86.92 &80.62 &84.49 \\
MetaCS &73.33 &75.34 &68.76 &75.76 &76.10 &72.09 &86.98 &80.06 &84.95 \\
MAMO &73.64 &75.48 &67.22 &77.27 &75.83 &72.85 &87.07 &80.44 &84.48 \\ \hline
TaNP (w/o tm) &75.45 &75.21 &69.06 &77.06 &76.19 &72.54 &87.42 &80.50 &84.59 \\
TaNP (FiLM) &76.06  &76.31 &$\bm{70.73}$ &77.92 &77.21 &$\bm{74.87}$ &88.33 &81.19 &85.04 \\ 
TaNP (Gating-FiLM) &$\bm{76.36}$ &$\bm{77.18}$ &70.12 &$\bm{79.00}$ &$\bm{77.30}$ &73.92 &$\bm{88.64}$
&$\bm{82.10}$ &$\bm{85.94}$ \\ \hline
\end{tabular}
}
\label{lastfm}
\end{table*}

\begin{table*}[]
\centering
\caption{Performance (\%) comparison of user cold-start recommendation on Gowalla.}
\resizebox{17.5cm}{!}{
\begin{tabular}{cccccccccc}
\hline
Model &P@5 &NDCG@5 &MAP@5 &P@7 &NDCG@7 &MAP@7 &P@10 &NDCG@10 &MAP@10 \\ \hline \hline
PPR &60.48 &62.92 &53.09 &61.89 &64.43 &56.46 &62.00  &66.44 &59.31 \\
NeuMF &55.98 &57.99 &51.02 &59.30 &60.36 &53.34 &60.80 &64.47 &56.17 \\
DropoutNet &62.06 &65.79 &55.46 &63.44 &66.30 &56.92 &64.73 &67.85 &60.10 \\
MetaLWA &64.67 &65.53 &55.99 &65.12 &66.53 &56.74 &69.09 &66.61 &60.35 \\
MetaNLBA &67.60 &67.93 &59.56 &68.38 &68.45 &61.07 &70.66 &69.45 &63.00 \\
MeLU &67.85 &69.40 &60.97 &70.14 &69.43 &63.66 &70.40 &72.13 &63.95 \\
MetaCS &66.14 &67.39 &58.69 &67.52 &67.82 &60.74 &69.29 &69.63 &61.87 \\
MAMO &67.97 &69.52 &61.07 &71.03 &70.54 &63.73 &71.43 &73.13 &64.99 \\ \hline
TaNP (w/o tm) &68.25 &69.76 &61.29 &70.83 &70.03 &64.15 &71.08 &72.72 &64.39 \\
TaNP (FiLM) &70.94  &71.14 &$\bm{64.35}$  &72.08 &$\bm{72.18}$ &65.95 &72.65 &74.12 &66.39 \\ 
TaNP (Gating-FiLM) &$\bm{71.43}$ &$\bm{71.60}$ &64.29 &$\bm{72.58}$ &71.99 &$\bm{66.16}$ &$\bm{73.11}$ &$\bm{74.57}$ &$\bm{66.79}$ \\ \hline
\end{tabular}
}

\label{gowalla}
\end{table*}

\subsubsection{Data Preprocessing}
For MovieLens-1M, we use the contents of users and items, i.e., side-information, which are collected by~\cite{lee2019melu}. It includes the following category contents of users: gender, age, occupation and zip code. The item contents have publication year, rate, genre, director and actor. To verify that our model is applicable in different experimental settings, we transform Last.FM and Gowalla into implicit feedback datasets. Following the data preprocessing in~\cite{wang2019multi}, we generate negative samples for the query sets in these two datasets. 
\par
For each dataset, the division ratio of training, validation and test sets is 7:1:2. As suggested by previous works~\cite{lee2019melu, Dong2020MAMOMM, lu2020meta}, we only keep the users whose item-consumption history length is between 40 and 200. To generalize well with only a few samples, we set the number of interactions in support set, e.g., $N_{S_i}$ as a small value ($N_{S_i}$=20/15/10), and remaining interactions are set as the query set. 
\subsection{Experimental Setting}
\subsubsection{Evaluation Metrics} 
To evaluate the recommendation performance, three frequently-used metrics: Precision~(P)@N, Normalized Discounted Cumulative Gain~(NDCG)@N and Mean Average Precision~(MAP)@N are adopted (N = 5, 7, 10). For each metric, we report the average results for all users in the test set. Following previous meta-learning recommenders~\cite{lee2019melu,bharadhwaj2019meta,lu2020meta,Dong2020MAMOMM}, we only predict the score of each item in the query set for each user. 
\begin{figure*}[t]
 \centering
 \subfigure[MovieLens-1M]{
  \includegraphics[width=5cm, height=4.4cm]{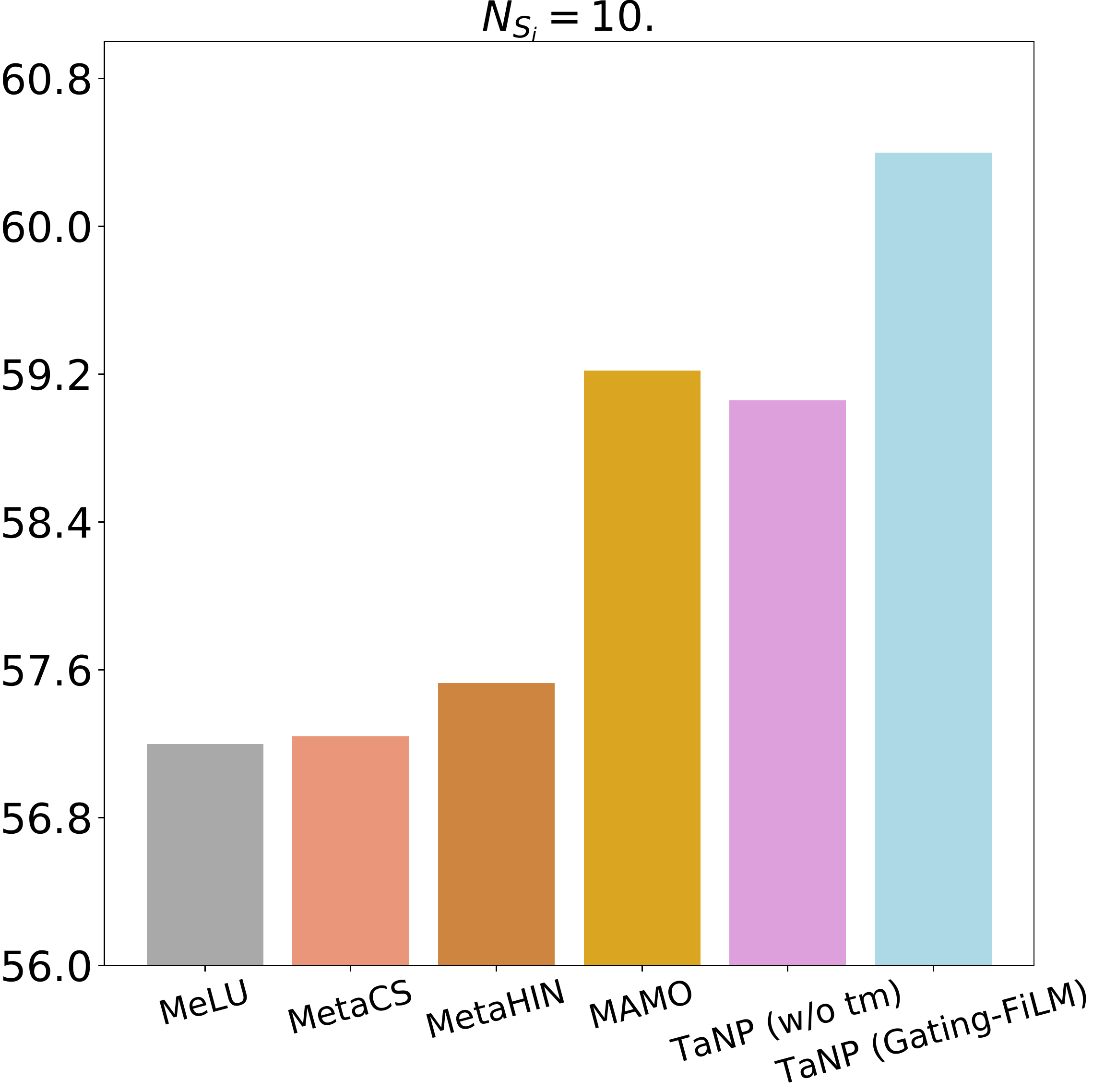}
 }
 \subfigure[Last.FM]{
  \includegraphics[width=5cm, height=4.4cm]{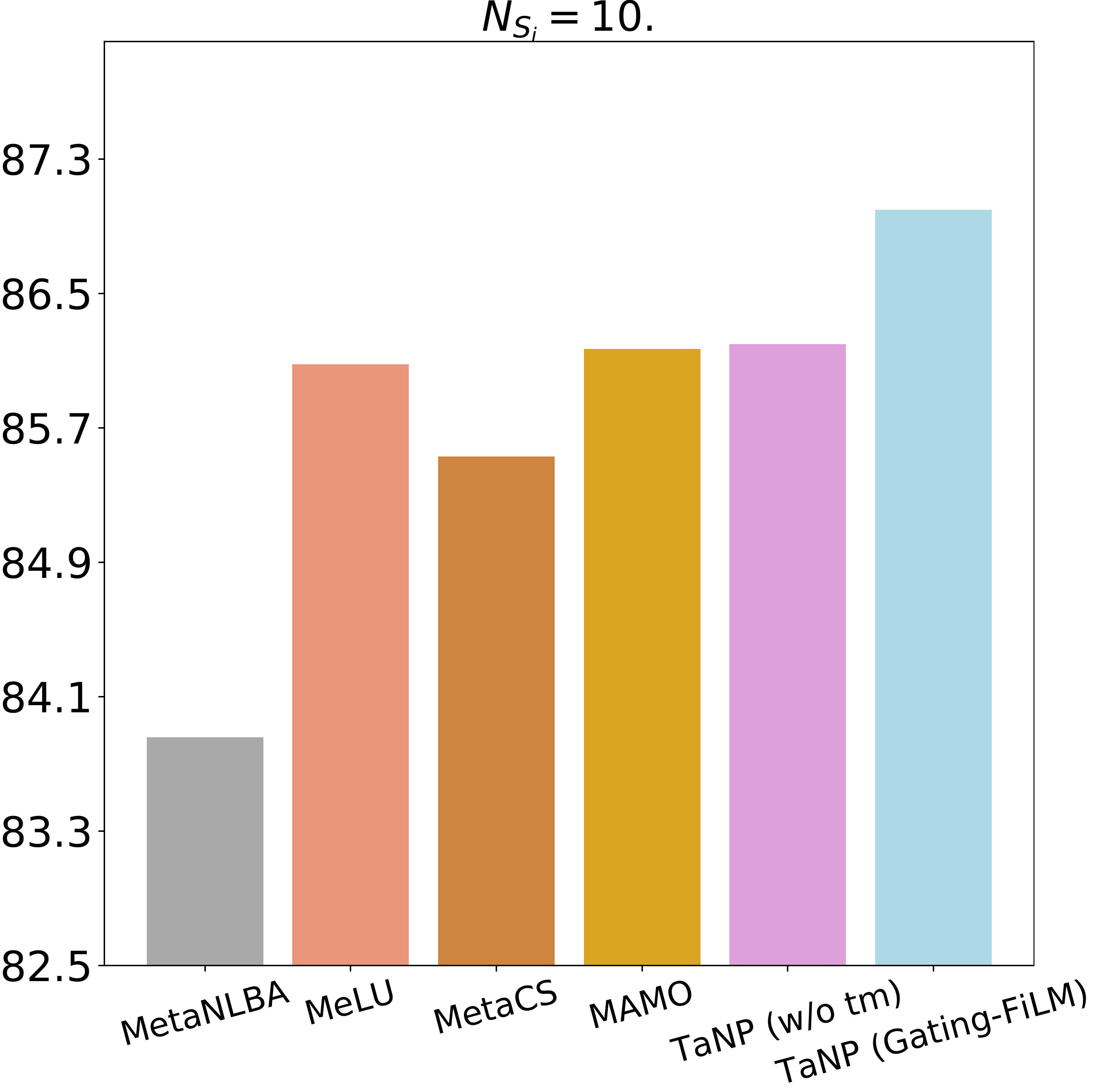}
 }
 \subfigure[Gowalla]{
  \includegraphics[width=5cm, height=4.4cm]{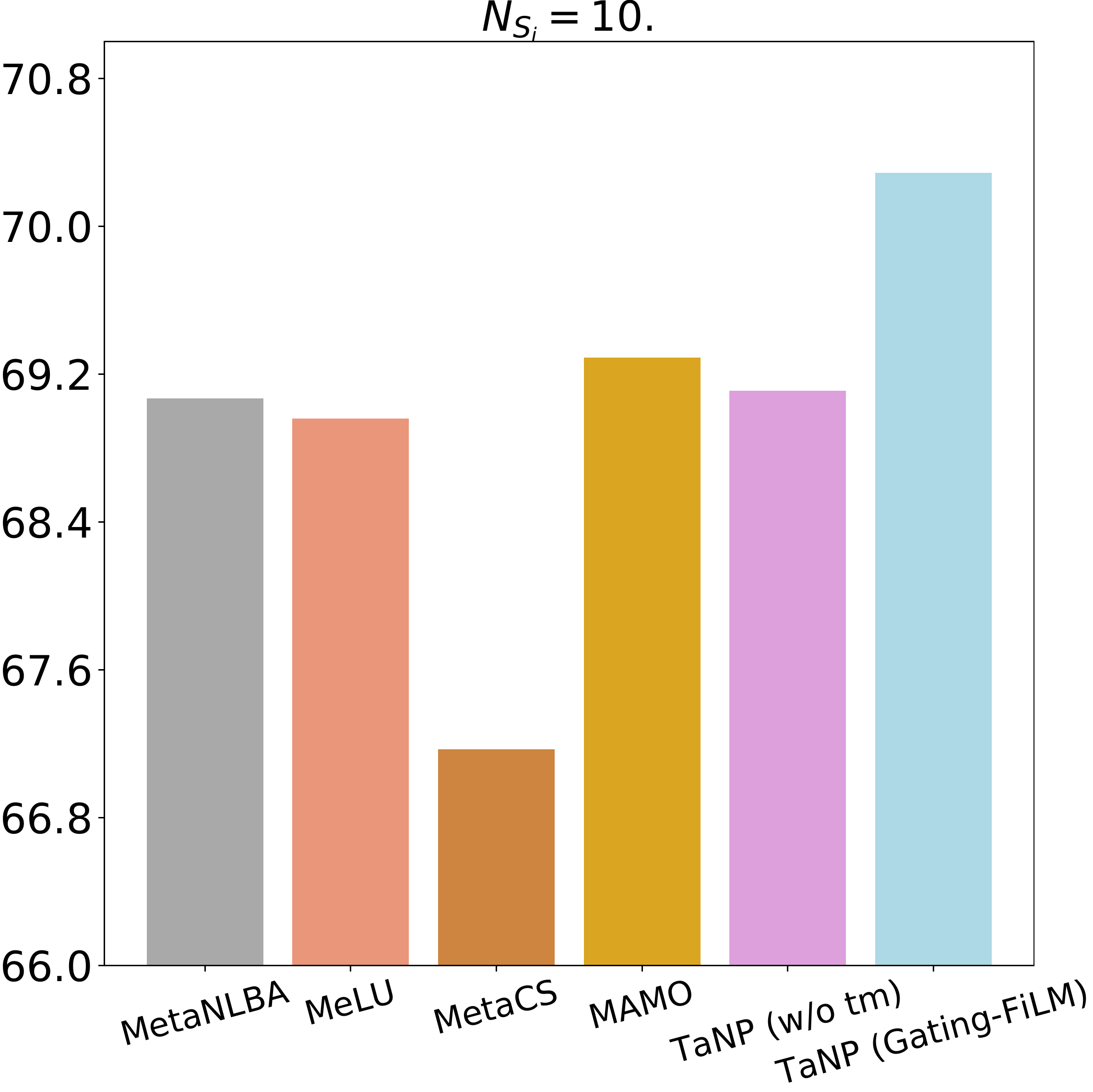}
 }
  \subfigure[MovieLens-1M]{
  \includegraphics[width=5cm, height=4.4cm]{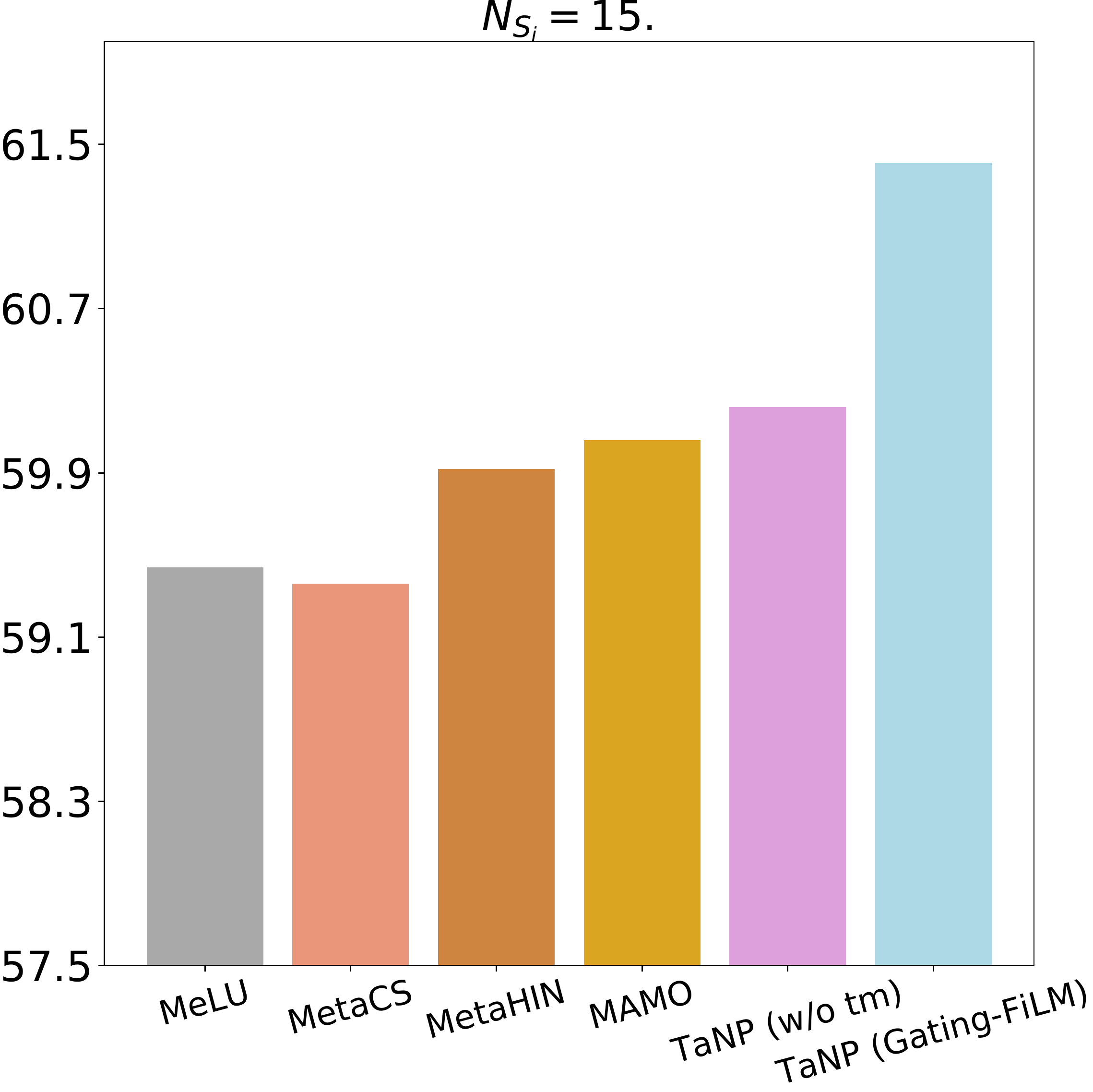}
 }
 \subfigure[Last.FM]{
  \includegraphics[width=5cm, height=4.4cm]{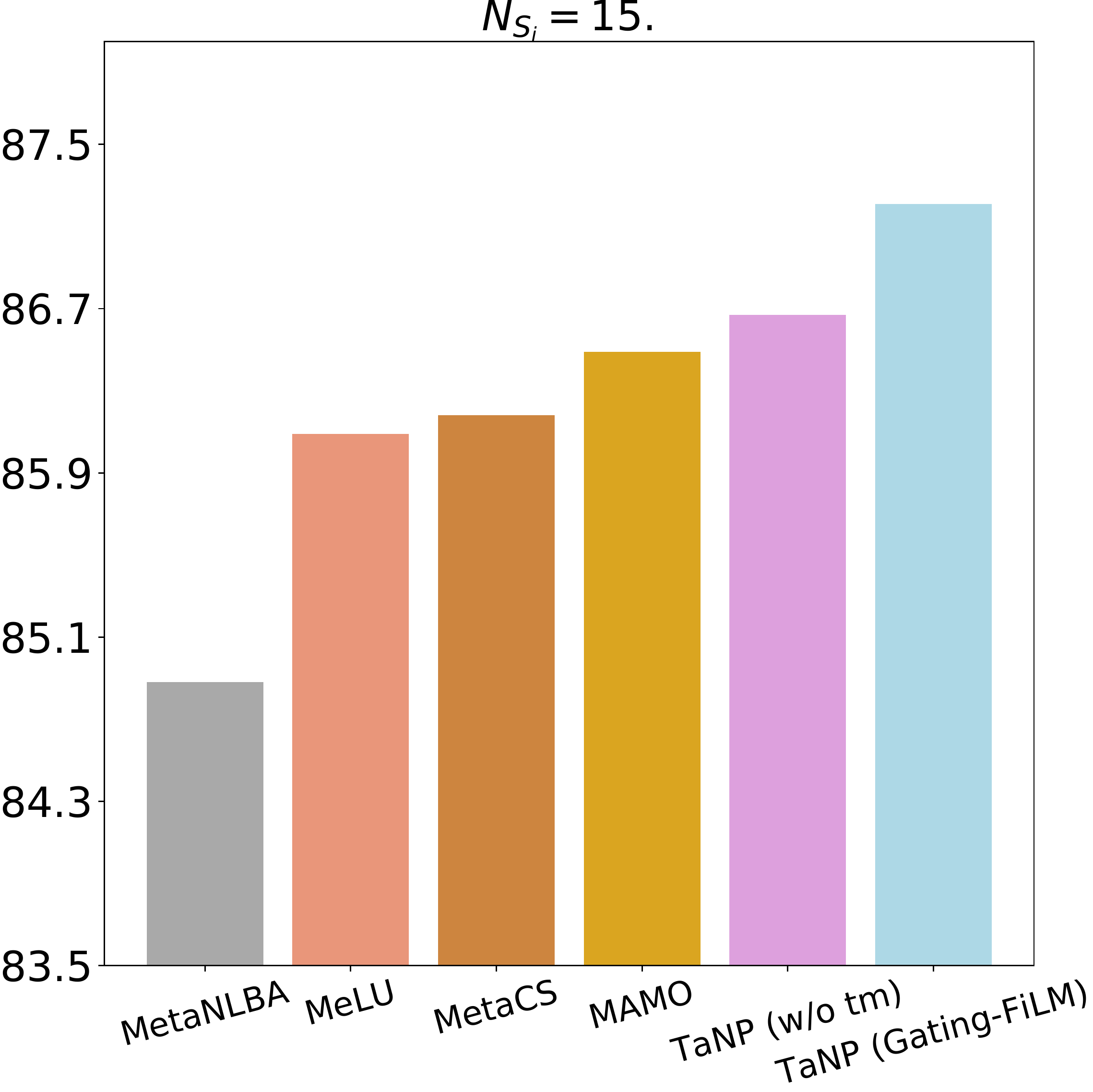}
 }
 \subfigure[Gowalla]{
  \includegraphics[width=5cm, height=4.4cm]{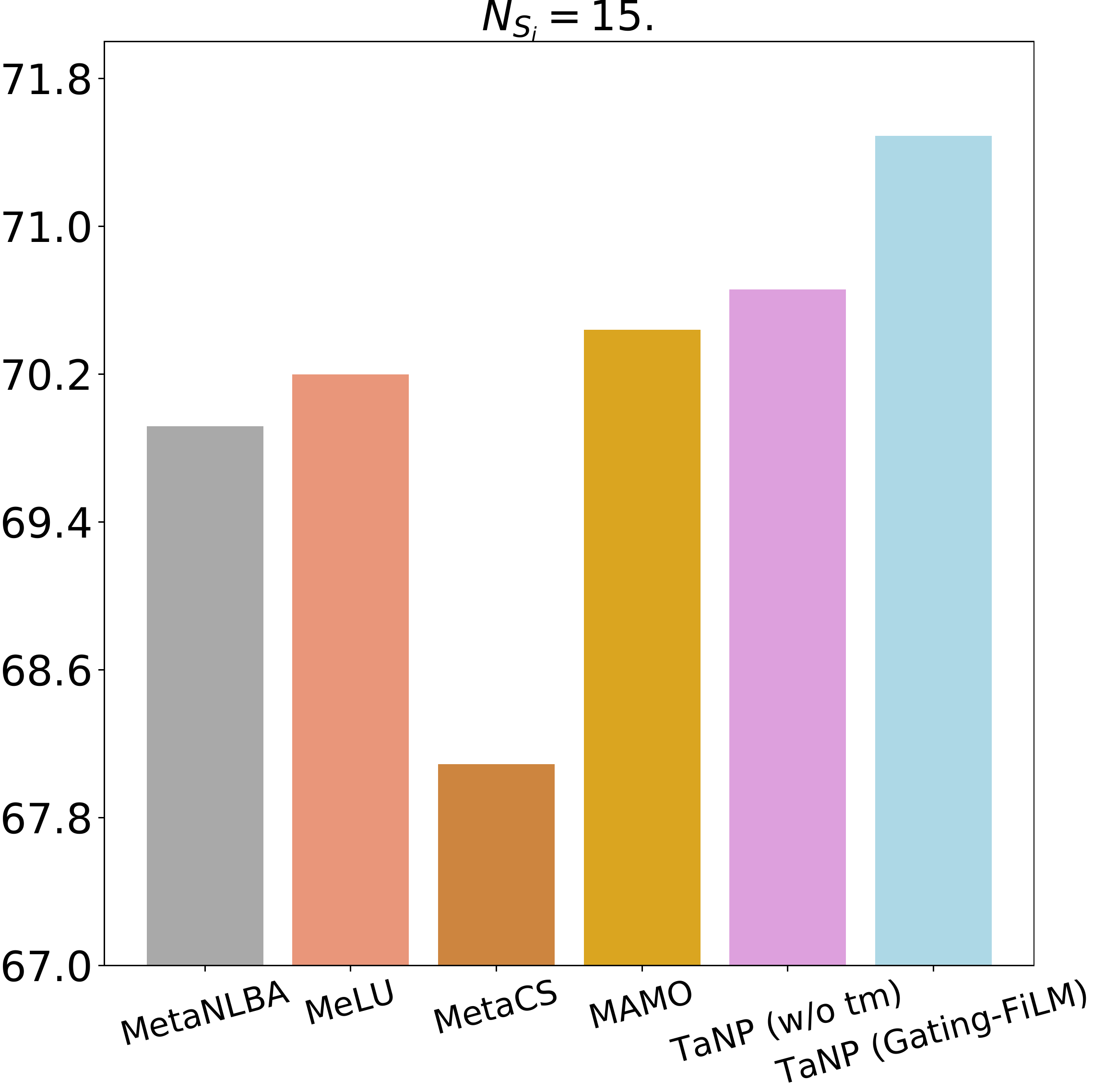}
 }
\caption{Performance (\%) comparison of user cold-start recommendation with different lengths of support set. The results of all datasets are provided. The first row shows the empirical results when $N_{S_i}=10$, and the second row shows the empirical results when $N_{S_i}=15$.
}
\label{sup_length}
\vspace{-3mm}
\end{figure*}
\subsubsection{Compared Models}
We compare our method with the following models: 
\begin{itemize}
    \item PPR~\cite{park2009pairwise} is a well-known CF-based method to solve cold-start problems. PPR first constructs the profiles for user-item pairs by the outer product over their individual features, then it develops a novel regression framework to estimate the pairwise user preferences. 
    \item NeuMF~\cite{He2017NeuralCF} is also a CF-based model that exploits DNNs to capture the non-linear feature interaction between user and item. NeuMF is a classic recommendation model, but it is not especially designed for cold-start problems. We adopt it here to test its effectiveness. 
    \item DropoutNet~\cite{volkovs2017dropoutnet} is a content-based model to solve cold-start problems. It applies the dropout mechanism to input during training to condition for missing user interactions.
    \item MetaLWA~\cite{vartak2017meta} and MetaNLBA~\cite{vartak2017meta} are the first meta-learning recommenders for cold-start problems. Different from our work, they focus on item cold-start recommendation. They generate two class-level prototype representations to learn item similarity. MetaLWA learns a linear classifier whose weights are determined by the user's interaction history, and MetaNLBA learns a DNN with the task-dependent bias for each layer. 
    \item MeLU~\cite{lee2019melu} handles cold-start problems by applying the framework of MAML. Based on the learned parameter initialization, MeLU can make recommendations for cold-start users via a few steps of gradient updates. 
    \item MetaCS~\cite{bharadhwaj2019meta} is similar to MeLU, which also exploits MAML to estimate the user preference. It includes three model variants, and we choose the best one according to their reported results. 
    \item MetaHIN~\cite{lu2020meta} combines MAML with HINs to alleviate cold-start problems from model and data levels. The rich semantic of HINs provides a finer-grained prior which is beneficial to fast adaptations of new tasks. 
    \item MAMO~\cite{Dong2020MAMOMM} is a memory-augmented framework of MAML. MAMO assumes that current MAML-based models often suffer from gradient degradation ending up with a local optima when handling users who show different gradient descent directions comparing with the majority of users in the training set. So it designs the task-specific and feature-specific memory matrices to solve this problem.
\end{itemize}
\subsubsection{Implementation Details}
The source codes of NeuMF~\footnote{https://github.com/hexiangnan/neural\_collaborative\_filtering}, DropoutNet~\footnote{https://github.com/layer6ai-labs/DropoutNet}, MeLU~\footnote{https://github.com/hoyeoplee/MeLU}, MetaHIN~\footnote{https://github.com/rootlu/MetaHIN} and MAMO~\footnote{https://github.com/dongmanqing/Code-for-MAMO} have been released, and we modify the parts of data input and evaluations to fit our experimental settings. Other baselines are reproduced by ourselves. MetaLWA and MetaNLBA are only applicable to implicit feedback datasets, so the results of them on MovieLens-1M are not provided. While DropoutNet is a content-based method, the used input dropout is a general technique that can be also employed on Last.FM and Gowalla. In addition, both MetaHIN and MAMO are closely connected with the side-information of user and items. In particular, the meta-paths used in MetaHIN are constructed according to the node types of users, thus only the results of MetaHIN for MovieLens-1M are reported. The attention calculation in MAMO is obtained via user contents and the proposed profile memory. We replace this part with the original memory mechanism enabling MAMO to be deployed on Last.FM and Gowalla. 
\par
To make fair comparisons, the dimension sizes of user and item embeddings are fixed as 32 in embedding layers. All models are fully iterated with 150 epochs for convergence. In our model, the number of stacked layers $l$ for all modules in our model ($h_{\theta}$, $m_{\phi}$ and $g_{\omega_{i}}$) is 3, the learning rate is 5e-5 and the degree of freedom $\alpha$ is 1.0. Other hyper-parameters are selected according to the performances on validation datasets. The hidden size of each layer is selected from \{8, 16, 32, 64, 128\}, the hyper-parameter $\lambda$ in Eq.(\ref{total_loss}) is selected from \{1.0, 0.5, 0.1, 0.05, 0.01\}, and the number of soft cluster controids $k$ in the global pool $\bm{A}$ ranges from 10 to 50 with the step length 10.
\subsection{Performance Comparison (RQ1)}
Table~\ref{movie},~\ref{lastfm} and~\ref{gowalla} demonstrate the performances of all models for user cold-start recommendation on MovieLens-1M, Last.FM and Gowalla. The size of support set $N_{S_i}$ is set as 20 in this section. The best performances are in bold. TaNP (w/o tm) denotes an ablation study of our model, in which the proposed task-adaptive mechanism is removed. TaNP (FiLM) and TaNP (Gating-FiLM) are two variants of our model using different modulation strategies. From these Tables, we can draw the following conclusions:
\vspace{-\topsep}
\begin{itemize}
    \item TaNP consistently yields the best performances on all datasets. Compared with state-of-the-art meta-learning recommenders, the most obvious improvements of TaNP are listed here: for the MovieLens-1M, TaNP brings 4.2\% improvements in terms of P@5; TaNP achieves 3.6\% result lifts in terms of P@5 on Last.FM; For the Gowalla, TaNP provides 5.4\% improvements in terms of MAP@5. 
    \item Compared with those meta-learning recommenders based on MAML, we find that TaNP (w/o tm) achieves competitive performances. It demonstrates that our NP framework is suitable to user cold-start recommendation.
    \item In contrast to TaNP (w/o tm), the improvements of our variants, i.e., TaNP (FiLM) and TaNP (Gating-FiLM) are significant, and TaNP (Gating-FiLM) is slightly better than TaNP (FiLM). It demonstrates that the effectiveness of our task-adaptive mechanism and the importance of learning the relevance of different tasks.
    \item MetaLWA, MetaNLBA and MetaHIN are constrained by dataset types. Different from them, TaNP is a more general meta-learning recommender which can be used in different experimental settings.
\end{itemize}
\subsection{Impact of $N_{S_i}$ (RQ2)}
We change the number of interactions in $S_i$ for three datasets to test the effectiveness of our methods. Concretely, $N_{S_i}$ is set as 10 and 15 and the evaluation metric is P@10. According to the above results in Section 5.3, we select the following strong baselines: MetaNLBA, MeLU, MetaCS, MetaHIN and MAMO. Here we only report the results of TaNP (Gating-FiLM), since it achieves the better performances compared with TaNP (FiLM). The empirical results are provided in Figure~\ref{sup_length}, from which we can draw the following conclusions: 
\begin{itemize}
    \item For all datasets, when the number of interactions in the support set has been reduced, TaNP (Gating-FiLM) still maintains the better performances compared with other meta-learning models.
    \item TaNP (Gating-FiLM) performs better than TaNP (w/o tm) in all experimental settings. Compared with MeLU and MetaCS, the improvements of MAMO are also obvious. Through these two comparisons, we find that it is essential to use an effective adaptation for handling different tasks.
    \item Our model is less influenced by the decrease of interactions. The possible reason is that TaNP applies a random function to the constructions of $\tau_i$ for modeling different function realisations. In contrast, other baselines, such as the MetaNLBA and MetaCS, are more sensitive to the change of $N_{S_i}$.
\end{itemize}

\begin{figure}
\begin{center}
\includegraphics[width=7cm,height=5cm]{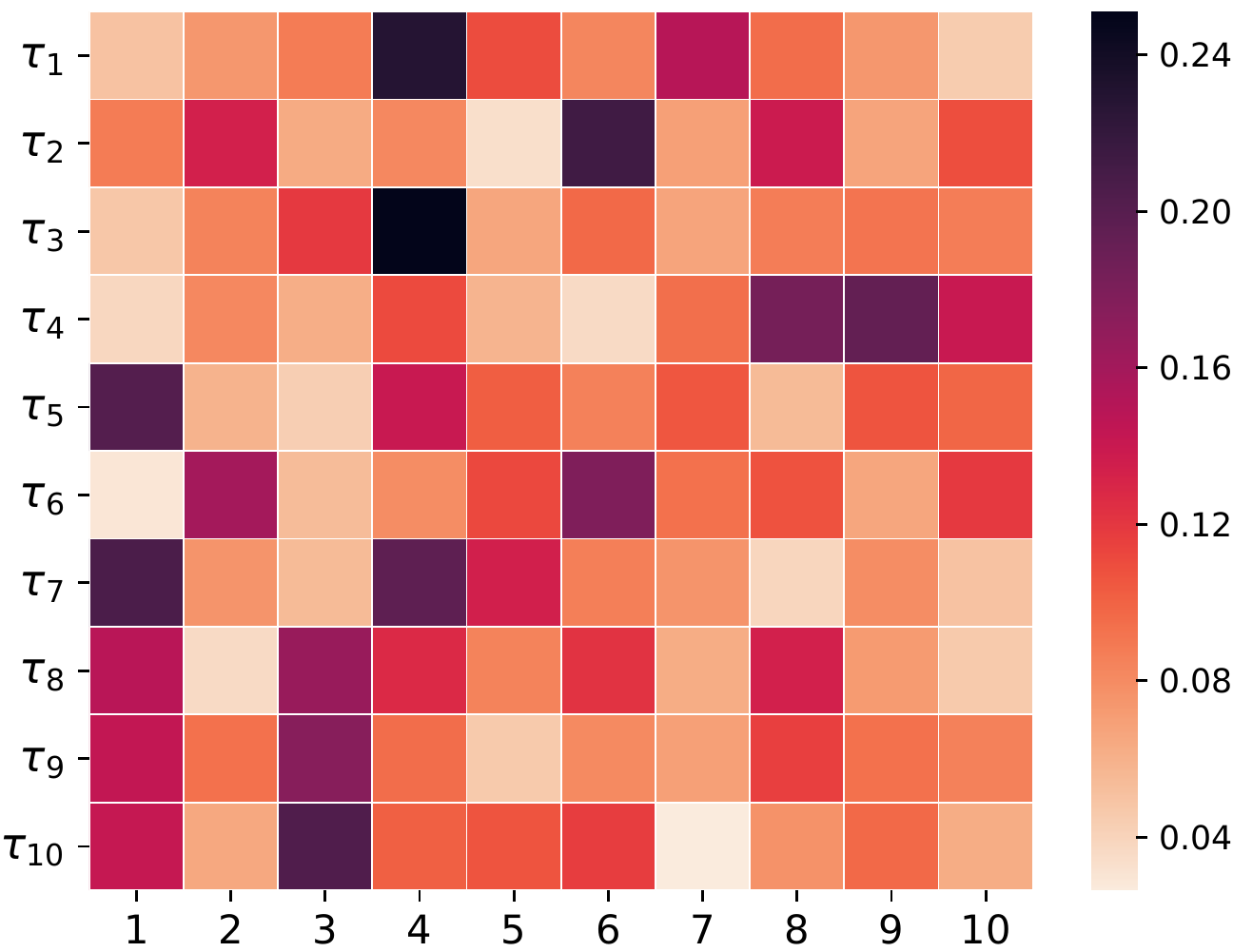}
\caption{Visualization of soft cluster assignments of 10 tasks $\{\tau_{i}\}_{i=1}^{10}$. X axis represents $k=10$ cluster centroids. If $\tau_{i}$ and $\tau_{j}$ have high scores (\textit{i.e.,} dark color) on the same cluster centroids, we assume they may share some similarities.}
\label{heatmap}
\end{center}
\vspace{-3mm}
\end{figure}

\subsection{Visual Analysis (RQ3)}
In our model, the proposed adaptive mechanism includes a global pool $\bm{A}$ that implicitly represents $k$ cluster centroids to capture the relevance of different tasks. Each task $\tau_i$ would interact with $\bm{A}$ to derive the corresponding soft cluster assignments, and we randomly pick 10 tasks from the training set of MovieLens-1M with their corresponding soft cluster assignments. 
The visual result is shown in Figure~\ref{heatmap}, and we have the following conclusions: 
\begin{itemize}
    \item Our model can capture the similarity of tasks. For example, the highest normalized scores of $\tau_5$ and $\tau_7$ are simultaneously assigned to the first and the fourth clusters. We infer that these two tasks are closely relevant. The side-information also provide some cues. Concretely, we find the corresponding user ids $u_5$ and $u_7$ are of the same gender, and they share two movie items in support sets. 
    \item The difference between tasks can be also well distinguished. For example, the soft assignments of $\tau_1$ and $\tau_4$ indicate that they are largely different.
\end{itemize}
\par
Overall, the task relevance including the task similarity and the task difference is learned by the customization module. Moreover, such relevance is incorporated into the final task embedding which further facilitates the task-related modulations of decoder parameters reliably.
\begin{figure}
\begin{center}
\includegraphics[width=8.5cm,height=5cm]{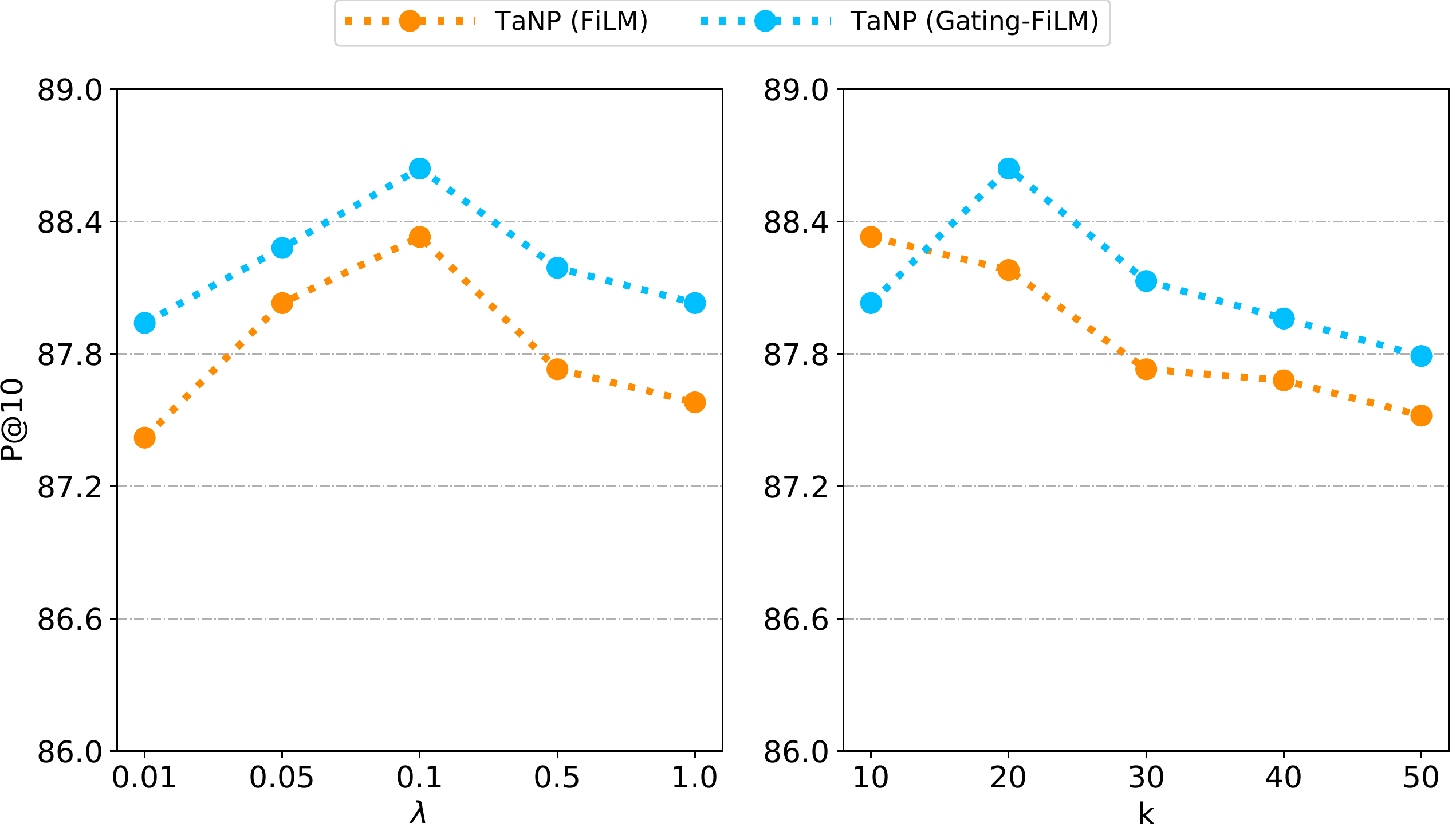}
\caption{Empirical results of parameter sensitivity.}
\label{hyper}
\end{center}
\vspace{-3mm}
\end{figure}

\subsection{Hyper-parameter Analysis (RQ4)}
In this section, we investigate the parameter sensitivity of our model with respect to two main hyper-parameters, i.e., $\lambda$ in the loss function and $k$ in $\bm{A}$. This experiment is conducted on Last.FM. As shown in Figure~\ref{hyper}, TaNP (FiLM) achieves the best result when $\lambda=0.1$ and $k=10$, and TaNP (Gating-FiLM) achieves the best result when $\lambda=0.1$ and $k=20$. Thus, choosing relative small values of $\lambda$ and $k$ is sensible. In addition, two variants of our model are robust to the changes of $\lambda$ and $k$. Even in the worst cases of  $\lambda$ and $k$, they are still better than other baselines shown in Table~\ref{lastfm}.


\section{Conclusion}
In this paper, we develop a novel meta-learning model called TaNP for user cold-start recommendation. TaNP is a generic framework which maps the observed interactions to the desired predictive distribution conditioned on a learned conditional prior. Furthermore, a novel task-adaptive mechanism is also introduced into TaNP for learning the relevance of different tasks as well as modulating the decoder parameters. Extensive results show that TaNP outperforms several state-of-the-art meta-learning recommenders consistently. 

\section*{Acknowledgement}
This work was supported in part by the NSFC (No. 61872360), the ARC DECRA (No. DE200100964), and the Youth Innovation Promotion Association CAS (No. 2017210).

\balance
\bibliographystyle{ACM-Reference-Format}
\bibliography{main}


\begin{thebibliography}{53}


\ifx \showCODEN    \undefined \def \showCODEN     #1{\unskip}     \fi
\ifx \showDOI      \undefined \def \showDOI       #1{#1}\fi
\ifx \showISBNx    \undefined \def \showISBNx     #1{\unskip}     \fi
\ifx \showISBNxiii \undefined \def \showISBNxiii  #1{\unskip}     \fi
\ifx \showISSN     \undefined \def \showISSN      #1{\unskip}     \fi
\ifx \showLCCN     \undefined \def \showLCCN      #1{\unskip}     \fi
\ifx \shownote     \undefined \def \shownote      #1{#1}          \fi
\ifx \showarticletitle \undefined \def \showarticletitle #1{#1}   \fi
\ifx \showURL      \undefined \def \showURL       {\relax}        \fi
\providecommand\bibfield[2]{#2}
\providecommand\bibinfo[2]{#2}
\providecommand\natexlab[1]{#1}
\providecommand\showeprint[2][]{arXiv:#2}

\bibitem[\protect\citeauthoryear{Antoniou, Edwards, and Storkey}{Antoniou
  et~al\mbox{.}}{2019}]%
        {antoniou2018how}
\bibfield{author}{\bibinfo{person}{Antreas Antoniou}, \bibinfo{person}{Harrison
  Edwards}, {and} \bibinfo{person}{Amos Storkey}.}
  \bibinfo{year}{2019}\natexlab{}.
\newblock \showarticletitle{How to train your {MAML}}. In
  \bibinfo{booktitle}{\emph{International Conference on Learning
  Representations (ICLR)}}.
\newblock


\bibitem[\protect\citeauthoryear{Bharadhwaj}{Bharadhwaj}{2019}]%
        {bharadhwaj2019meta}
\bibfield{author}{\bibinfo{person}{Homanga Bharadhwaj}.}
  \bibinfo{year}{2019}\natexlab{}.
\newblock \showarticletitle{Meta-Learning for User Cold-Start Recommendation}.
  In \bibinfo{booktitle}{\emph{2019 International Joint Conference on Neural
  Networks (IJCNN)}}.
\newblock


\bibitem[\protect\citeauthoryear{Bianchi, Cesaro, Ciceri, Dagrada, Gasparin,
  Grattarola, Inajjar, Metelli, and Cella}{Bianchi et~al\mbox{.}}{2017}]%
        {bianchi2017content}
\bibfield{author}{\bibinfo{person}{Mattia Bianchi}, \bibinfo{person}{Federico
  Cesaro}, \bibinfo{person}{Filippo Ciceri}, \bibinfo{person}{Mattia Dagrada},
  \bibinfo{person}{Alberto Gasparin}, \bibinfo{person}{Daniele Grattarola},
  \bibinfo{person}{Ilyas Inajjar}, \bibinfo{person}{Alberto~Maria Metelli},
  {and} \bibinfo{person}{Leonardo Cella}.} \bibinfo{year}{2017}\natexlab{}.
\newblock \showarticletitle{Content-based approaches for cold-start job
  recommendations}. In \bibinfo{booktitle}{\emph{Proceedings of the Recommender
  Systems Challenge}}.
\newblock


\bibitem[\protect\citeauthoryear{Blei, Ng, and Jordan}{Blei
  et~al\mbox{.}}{2003}]%
        {blei2003latent}
\bibfield{author}{\bibinfo{person}{David~M Blei}, \bibinfo{person}{Andrew~Y
  Ng}, {and} \bibinfo{person}{Michael~I Jordan}.}
  \bibinfo{year}{2003}\natexlab{}.
\newblock \showarticletitle{Latent dirichlet allocation}.
\newblock \bibinfo{journal}{\emph{Journal of machine Learning Research (JMLR)}}
  (\bibinfo{year}{2003}).
\newblock


\bibitem[\protect\citeauthoryear{Brockschmidt}{Brockschmidt}{2020}]%
        {brockschmidt2019gnn}
\bibfield{author}{\bibinfo{person}{Marc Brockschmidt}.}
  \bibinfo{year}{2020}\natexlab{}.
\newblock \showarticletitle{Gnn-film: Graph neural networks with feature-wise
  linear modulation}. In \bibinfo{booktitle}{\emph{International Conference on
  Machine Learning (ICML)}}.
\newblock


\bibitem[\protect\citeauthoryear{Cadene, Ben-Younes, Cord, and Thome}{Cadene
  et~al\mbox{.}}{2019}]%
        {cadene2019murel}
\bibfield{author}{\bibinfo{person}{Remi Cadene}, \bibinfo{person}{Hedi
  Ben-Younes}, \bibinfo{person}{Matthieu Cord}, {and} \bibinfo{person}{Nicolas
  Thome}.} \bibinfo{year}{2019}\natexlab{}.
\newblock \showarticletitle{Murel: Multimodal relational reasoning for visual
  question answering}. In \bibinfo{booktitle}{\emph{IEEE Conference on Computer
  Vision and Pattern Recognition (CVPR)}}.
\newblock


\bibitem[\protect\citeauthoryear{Cao*, Lin*, Guo, Liu, Liu, and Wang}{Cao*
  et~al\mbox{.}}{2021}]%
        {bigi2021}
\bibfield{author}{\bibinfo{person}{Jiangxia Cao*}, \bibinfo{person}{Xixun
  Lin*}, \bibinfo{person}{Shu Guo}, \bibinfo{person}{Luchen Liu},
  \bibinfo{person}{Tingwen Liu}, {and} \bibinfo{person}{Bin Wang}.}
  \bibinfo{year}{2021}\natexlab{}.
\newblock \showarticletitle{Bipartite Graph Embedding via Mutual Information
  Maximization}. In \bibinfo{booktitle}{\emph{ACM International Conference on
  Web Search and Data Mining (WSDM)}}.
\newblock


\bibitem[\protect\citeauthoryear{Das, Datar, Garg, and Rajaram}{Das
  et~al\mbox{.}}{2007}]%
        {das2007google}
\bibfield{author}{\bibinfo{person}{Abhinandan~S Das}, \bibinfo{person}{Mayur
  Datar}, \bibinfo{person}{Ashutosh Garg}, {and} \bibinfo{person}{Shyam
  Rajaram}.} \bibinfo{year}{2007}\natexlab{}.
\newblock \showarticletitle{Google news personalization: scalable online
  collaborative filtering}. In \bibinfo{booktitle}{\emph{The World Wide Web
  Conference (WWW)}}.
\newblock


\bibitem[\protect\citeauthoryear{Dong, Yuan, Yao, Xu, and Zhu}{Dong
  et~al\mbox{.}}{2020}]%
        {Dong2020MAMOMM}
\bibfield{author}{\bibinfo{person}{Manqing Dong}, \bibinfo{person}{Feng Yuan},
  \bibinfo{person}{Lina Yao}, \bibinfo{person}{Xiwei Xu}, {and}
  \bibinfo{person}{Liming Zhu}.} \bibinfo{year}{2020}\natexlab{}.
\newblock \showarticletitle{MAMO: Memory-Augmented Meta-Optimization for
  Cold-start Recommendation}. In \bibinfo{booktitle}{\emph{ACM Knowledge
  Discovery and Data Mining (KDD)}}.
\newblock


\bibitem[\protect\citeauthoryear{Du, Wang, Yang, Zhou, and Tang}{Du
  et~al\mbox{.}}{2019}]%
        {du2019scenariometa}
\bibfield{author}{\bibinfo{person}{Zhengxiao Du}, \bibinfo{person}{Xiaowei
  Wang}, \bibinfo{person}{Hongxia Yang}, \bibinfo{person}{Jingren Zhou}, {and}
  \bibinfo{person}{Jie Tang}.} \bibinfo{year}{2019}\natexlab{}.
\newblock \showarticletitle{Sequential Scenario-Specific Meta Learner for
  Online Recommendation}. In \bibinfo{booktitle}{\emph{ACM Knowledge Discovery
  and Data Mining (KDD)}}.
\newblock


\bibitem[\protect\citeauthoryear{Finn, Abbeel, and Levine}{Finn
  et~al\mbox{.}}{2017}]%
        {finn2017model}
\bibfield{author}{\bibinfo{person}{Chelsea Finn}, \bibinfo{person}{Pieter
  Abbeel}, {and} \bibinfo{person}{Sergey Levine}.}
  \bibinfo{year}{2017}\natexlab{}.
\newblock \showarticletitle{Model-agnostic meta-learning for fast adaptation of
  deep networks}. In \bibinfo{booktitle}{\emph{International Conference on
  Machine Learning (ICML)}}.
\newblock


\bibitem[\protect\citeauthoryear{Gao, Yang, Wu, Zhou, Lu, and Hu}{Gao
  et~al\mbox{.}}{2018}]%
        {gao2018recommendation}
\bibfield{author}{\bibinfo{person}{Li Gao}, \bibinfo{person}{Hong Yang},
  \bibinfo{person}{Jia Wu}, \bibinfo{person}{Chuan Zhou},
  \bibinfo{person}{Weixue Lu}, {and} \bibinfo{person}{Yue Hu}.}
  \bibinfo{year}{2018}\natexlab{}.
\newblock \showarticletitle{Recommendation with multi-source heterogeneous
  information}. In \bibinfo{booktitle}{\emph{International Joint Conference on
  Artificial Intelligence (IJCAI)}}.
\newblock


\bibitem[\protect\citeauthoryear{Garnelo, Rosenbaum, Maddison, Ramalho, Saxton,
  Shanahan, Teh, Rezende, and Eslami}{Garnelo et~al\mbox{.}}{2018a}]%
        {garnelo2018conditional}
\bibfield{author}{\bibinfo{person}{Marta Garnelo}, \bibinfo{person}{Dan
  Rosenbaum}, \bibinfo{person}{Chris~J Maddison}, \bibinfo{person}{Tiago
  Ramalho}, \bibinfo{person}{David Saxton}, \bibinfo{person}{Murray Shanahan},
  \bibinfo{person}{Yee~Whye Teh}, \bibinfo{person}{Danilo~J Rezende}, {and}
  \bibinfo{person}{SM Eslami}.} \bibinfo{year}{2018}\natexlab{a}.
\newblock \showarticletitle{Conditional neural processes}. In
  \bibinfo{booktitle}{\emph{International Conference on Machine Learning
  (ICML)}}.
\newblock


\bibitem[\protect\citeauthoryear{Garnelo, Schwarz, Rosenbaum, Viola, Rezende,
  Eslami, and Teh}{Garnelo et~al\mbox{.}}{2018b}]%
        {Garnelo2018NeuralP}
\bibfield{author}{\bibinfo{person}{Marta Garnelo}, \bibinfo{person}{Jonathan
  Schwarz}, \bibinfo{person}{Dan Rosenbaum}, \bibinfo{person}{Fabio Viola},
  \bibinfo{person}{Danilo~Jimenez Rezende}, \bibinfo{person}{S.~M.~Ali Eslami},
  {and} \bibinfo{person}{Yee~Whye Teh}.} \bibinfo{year}{2018}\natexlab{b}.
\newblock \showarticletitle{Neural Processes}. In
  \bibinfo{booktitle}{\emph{International Conference on Machine Learning
  (ICML)}}.
\newblock


\bibitem[\protect\citeauthoryear{Gordon, Bruinsma, Foong, Requeima, Dubois, and
  Turner}{Gordon et~al\mbox{.}}{[n.d.]}]%
        {gordon2019convolutional}
\bibfield{author}{\bibinfo{person}{Jonathan Gordon}, \bibinfo{person}{Wessel~P
  Bruinsma}, \bibinfo{person}{Andrew~YK Foong}, \bibinfo{person}{James
  Requeima}, \bibinfo{person}{Yann Dubois}, {and} \bibinfo{person}{Richard~E
  Turner}.} \bibinfo{year}{[n.d.]}\natexlab{}.
\newblock \showarticletitle{Convolutional conditional neural processes}. In
  \bibinfo{booktitle}{\emph{International Conference on Learning
  Representations (ICLR)}}.
\newblock


\bibitem[\protect\citeauthoryear{Gu, Zhou, and Ding}{Gu et~al\mbox{.}}{2010}]%
        {gu2010collaborative}
\bibfield{author}{\bibinfo{person}{Quanquan Gu}, \bibinfo{person}{Jie Zhou},
  {and} \bibinfo{person}{Chris Ding}.} \bibinfo{year}{2010}\natexlab{}.
\newblock \showarticletitle{Collaborative filtering: Weighted nonnegative
  matrix factorization incorporating user and item graphs}. In
  \bibinfo{booktitle}{\emph{SIAM International Conference on Data Mining
  (SDM)}}.
\newblock


\bibitem[\protect\citeauthoryear{He, Liao, Zhang, Nie, Hu, and Chua}{He
  et~al\mbox{.}}{2017}]%
        {He2017NeuralCF}
\bibfield{author}{\bibinfo{person}{Xiangnan He}, \bibinfo{person}{Lizi Liao},
  \bibinfo{person}{Hanwang Zhang}, \bibinfo{person}{Liqiang Nie},
  \bibinfo{person}{Xia Hu}, {and} \bibinfo{person}{Tat-Seng Chua}.}
  \bibinfo{year}{2017}\natexlab{}.
\newblock \showarticletitle{Neural Collaborative Filtering}. In
  \bibinfo{booktitle}{\emph{The World Wide Web Conference (WWW)}}.
\newblock


\bibitem[\protect\citeauthoryear{He, Zhang, Kan, and Chua}{He
  et~al\mbox{.}}{2016}]%
        {he2016fast}
\bibfield{author}{\bibinfo{person}{Xiangnan He}, \bibinfo{person}{Hanwang
  Zhang}, \bibinfo{person}{Min-Yen Kan}, {and} \bibinfo{person}{Tat-Seng
  Chua}.} \bibinfo{year}{2016}\natexlab{}.
\newblock \showarticletitle{Fast matrix factorization for online recommendation
  with implicit feedback}. In \bibinfo{booktitle}{\emph{ACM International
  Conference on Research on Development in Information Retrieval (SIGIR)}}.
\newblock


\bibitem[\protect\citeauthoryear{Kaiser, Nachum, Roy, and Bengio}{Kaiser
  et~al\mbox{.}}{2017}]%
        {kaiser2017learning}
\bibfield{author}{\bibinfo{person}{{\L}ukasz Kaiser}, \bibinfo{person}{Ofir
  Nachum}, \bibinfo{person}{Aurko Roy}, {and} \bibinfo{person}{Samy Bengio}.}
  \bibinfo{year}{2017}\natexlab{}.
\newblock \showarticletitle{Learning to remember rare events}. In
  \bibinfo{booktitle}{\emph{International Conference on Learning
  Representations (ICLR)}}.
\newblock


\bibitem[\protect\citeauthoryear{Kim, Mnih, Schwarz, Garnelo, Eslami,
  Rosenbaum, Vinyals, and Teh}{Kim et~al\mbox{.}}{2019}]%
        {Kim2019AttentiveNP}
\bibfield{author}{\bibinfo{person}{Hyunjik Kim}, \bibinfo{person}{Andriy Mnih},
  \bibinfo{person}{Jonathan Schwarz}, \bibinfo{person}{Marta Garnelo},
  \bibinfo{person}{S.~M.~Ali Eslami}, \bibinfo{person}{Dan Rosenbaum},
  \bibinfo{person}{Oriol Vinyals}, {and} \bibinfo{person}{Yee~Whye Teh}.}
  \bibinfo{year}{2019}\natexlab{}.
\newblock \showarticletitle{Attentive Neural Processes}. In
  \bibinfo{booktitle}{\emph{International Conference on Learning
  Representations (ICLR)}}.
\newblock


\bibitem[\protect\citeauthoryear{Kingma and Ba}{Kingma and Ba}{2015}]%
        {Kingma2015AdamAM}
\bibfield{author}{\bibinfo{person}{Diederik~P. Kingma} {and}
  \bibinfo{person}{Jimmy Ba}.} \bibinfo{year}{2015}\natexlab{}.
\newblock \showarticletitle{Adam: A Method for Stochastic Optimization}. In
  \bibinfo{booktitle}{\emph{International Conference on Learning
  Representations (ICLR)}}.
\newblock


\bibitem[\protect\citeauthoryear{Kingma and Welling}{Kingma and
  Welling}{2014}]%
        {kingma2013auto}
\bibfield{author}{\bibinfo{person}{Diederik~P Kingma} {and}
  \bibinfo{person}{Max Welling}.} \bibinfo{year}{2014}\natexlab{}.
\newblock \showarticletitle{Auto-encoding variational bayes}. In
  \bibinfo{booktitle}{\emph{International Conference on Learning
  Representations (ICLR)}}.
\newblock


\bibitem[\protect\citeauthoryear{Koren, Bell, and Volinsky}{Koren
  et~al\mbox{.}}{2009}]%
        {koren2009matrix}
\bibfield{author}{\bibinfo{person}{Yehuda Koren}, \bibinfo{person}{Robert
  Bell}, {and} \bibinfo{person}{Chris Volinsky}.}
  \bibinfo{year}{2009}\natexlab{}.
\newblock \showarticletitle{Matrix factorization techniques for recommender
  systems}.
\newblock \bibinfo{journal}{\emph{Computer}} (\bibinfo{year}{2009}).
\newblock


\bibitem[\protect\citeauthoryear{Kouki, Fakhraei, Foulds, Eirinaki, and
  Getoor}{Kouki et~al\mbox{.}}{2015}]%
        {kouki2015hyper}
\bibfield{author}{\bibinfo{person}{Pigi Kouki}, \bibinfo{person}{Shobeir
  Fakhraei}, \bibinfo{person}{James Foulds}, \bibinfo{person}{Magdalini
  Eirinaki}, {and} \bibinfo{person}{Lise Getoor}.}
  \bibinfo{year}{2015}\natexlab{}.
\newblock \showarticletitle{Hyper: A flexible and extensible probabilistic
  framework for hybrid recommender systems}. In \bibinfo{booktitle}{\emph{ACM
  Conference on Recommender Systems (RecSys)}}.
\newblock


\bibitem[\protect\citeauthoryear{Lee, Im, Jang, Cho, and Chung}{Lee
  et~al\mbox{.}}{2019}]%
        {lee2019melu}
\bibfield{author}{\bibinfo{person}{Hoyeop Lee}, \bibinfo{person}{Jinbae Im},
  \bibinfo{person}{Seongwon Jang}, \bibinfo{person}{Hyunsouk Cho}, {and}
  \bibinfo{person}{Sehee Chung}.} \bibinfo{year}{2019}\natexlab{}.
\newblock \showarticletitle{MeLU: Meta-Learned User Preference Estimator for
  Cold-Start Recommendation}. In \bibinfo{booktitle}{\emph{ACM Knowledge
  Discovery and Data Mining (KDD)}}.
\newblock


\bibitem[\protect\citeauthoryear{Li, Zhou, Chen, and Li}{Li
  et~al\mbox{.}}{2017}]%
        {li2017meta}
\bibfield{author}{\bibinfo{person}{Zhenguo Li}, \bibinfo{person}{Fengwei Zhou},
  \bibinfo{person}{Fei Chen}, {and} \bibinfo{person}{Hang Li}.}
  \bibinfo{year}{2017}\natexlab{}.
\newblock \showarticletitle{Meta-sgd: Learning to learn quickly for few-shot
  learning}.
\newblock \bibinfo{journal}{\emph{arXiv}} (\bibinfo{year}{2017}).
\newblock


\bibitem[\protect\citeauthoryear{Liu, Zhou, Wu, Hu, and Guo}{Liu
  et~al\mbox{.}}{2018}]%
        {liu2018social}
\bibfield{author}{\bibinfo{person}{Chunyi Liu}, \bibinfo{person}{Chuan Zhou},
  \bibinfo{person}{Jia Wu}, \bibinfo{person}{Yue Hu}, {and} \bibinfo{person}{Li
  Guo}.} \bibinfo{year}{2018}\natexlab{}.
\newblock \showarticletitle{Social recommendation with an essential preference
  space}. In \bibinfo{booktitle}{\emph{AAAI Conference on Artificial
  Intelligence (AAAI)}}.
\newblock


\bibitem[\protect\citeauthoryear{Lops, De~Gemmis, and Semeraro}{Lops
  et~al\mbox{.}}{2011}]%
        {lops2011content}
\bibfield{author}{\bibinfo{person}{Pasquale Lops}, \bibinfo{person}{Marco
  De~Gemmis}, {and} \bibinfo{person}{Giovanni Semeraro}.}
  \bibinfo{year}{2011}\natexlab{}.
\newblock \showarticletitle{Content-based recommender systems: State of the art
  and trends}.
\newblock In \bibinfo{booktitle}{\emph{Recommender Systems Handbook}}.
\newblock


\bibitem[\protect\citeauthoryear{Louizos, Shi, Schutte, and Welling}{Louizos
  et~al\mbox{.}}{2019}]%
        {NIPS2019_9079}
\bibfield{author}{\bibinfo{person}{Christos Louizos}, \bibinfo{person}{Xiahan
  Shi}, \bibinfo{person}{Klamer Schutte}, {and} \bibinfo{person}{Max Welling}.}
  \bibinfo{year}{2019}\natexlab{}.
\newblock \showarticletitle{The Functional Neural Process}. In
  \bibinfo{booktitle}{\emph{Annual Conference on Neural Information Processing
  Systems (NeurIPS)}}.
\newblock


\bibitem[\protect\citeauthoryear{Lu, Fang, and Shi}{Lu et~al\mbox{.}}{2020}]%
        {lu2020meta}
\bibfield{author}{\bibinfo{person}{Yuanfu Lu}, \bibinfo{person}{Yuan Fang},
  {and} \bibinfo{person}{Chuan Shi}.} \bibinfo{year}{2020}\natexlab{}.
\newblock \showarticletitle{Meta-learning on heterogeneous information networks
  for cold-start recommendation}. In \bibinfo{booktitle}{\emph{ACM Knowledge
  Discovery and Data Mining (KDD)}}.
\newblock


\bibitem[\protect\citeauthoryear{Ma, Zhou, Cui, Yang, and Zhu}{Ma
  et~al\mbox{.}}{2019}]%
        {ma2019learning}
\bibfield{author}{\bibinfo{person}{Jianxin Ma}, \bibinfo{person}{Chang Zhou},
  \bibinfo{person}{Peng Cui}, \bibinfo{person}{Hongxia Yang}, {and}
  \bibinfo{person}{Wenwu Zhu}.} \bibinfo{year}{2019}\natexlab{}.
\newblock \showarticletitle{Learning disentangled representations for
  recommendation}. In \bibinfo{booktitle}{\emph{Advances in Neural Information
  Processing Systems (NeurIPS)}}.
\newblock


\bibitem[\protect\citeauthoryear{Ma, Zhou, Yang, Cui, Wang, and Zhu}{Ma
  et~al\mbox{.}}{2020}]%
        {ma2020disentangled}
\bibfield{author}{\bibinfo{person}{Jianxin Ma}, \bibinfo{person}{Chang Zhou},
  \bibinfo{person}{Hongxia Yang}, \bibinfo{person}{Peng Cui},
  \bibinfo{person}{Xin Wang}, {and} \bibinfo{person}{Wenwu Zhu}.}
  \bibinfo{year}{2020}\natexlab{}.
\newblock \showarticletitle{Disentangled Self-Supervision in Sequential
  Recommenders}. In \bibinfo{booktitle}{\emph{ACM Knowledge Discovery and Data
  Mining (KDD)}}.
\newblock


\bibitem[\protect\citeauthoryear{Mnih and Gregor}{Mnih and Gregor}{2014}]%
        {mnih2014neural}
\bibfield{author}{\bibinfo{person}{Andriy Mnih} {and} \bibinfo{person}{Karol
  Gregor}.} \bibinfo{year}{2014}\natexlab{}.
\newblock \showarticletitle{Neural variational inference and learning in belief
  networks}. In \bibinfo{booktitle}{\emph{International Conference on Machine
  Learning (ICML)}}.
\newblock


\bibitem[\protect\citeauthoryear{Nichol and Schulman}{Nichol and
  Schulman}{2018}]%
        {nichol2018reptile}
\bibfield{author}{\bibinfo{person}{Alex Nichol} {and} \bibinfo{person}{John
  Schulman}.} \bibinfo{year}{2018}\natexlab{}.
\newblock \showarticletitle{Reptile: a scalable metalearning algorithm}.
\newblock \bibinfo{journal}{\emph{arXiv}} (\bibinfo{year}{2018}).
\newblock


\bibitem[\protect\citeauthoryear{Oksendal}{Oksendal}{2013}]%
        {oksendal2013stochastic}
\bibfield{author}{\bibinfo{person}{Bernt Oksendal}.}
  \bibinfo{year}{2013}\natexlab{}.
\newblock \bibinfo{booktitle}{\emph{Stochastic differential equations: an
  introduction with applications}}.
\newblock \bibinfo{publisher}{Springer Science \& Business Media}.
\newblock


\bibitem[\protect\citeauthoryear{Park and Chu}{Park and Chu}{2009}]%
        {park2009pairwise}
\bibfield{author}{\bibinfo{person}{Seung-Taek Park} {and} \bibinfo{person}{Wei
  Chu}.} \bibinfo{year}{2009}\natexlab{}.
\newblock \showarticletitle{Pairwise preference regression for cold-start
  recommendation}. In \bibinfo{booktitle}{\emph{ACM conference on Recommender
  systems (RecSys)}}.
\newblock


\bibitem[\protect\citeauthoryear{Perez, Strub, De~Vries, Dumoulin, and
  Courville}{Perez et~al\mbox{.}}{2018}]%
        {perez2018film}
\bibfield{author}{\bibinfo{person}{Ethan Perez}, \bibinfo{person}{Florian
  Strub}, \bibinfo{person}{Harm De~Vries}, \bibinfo{person}{Vincent Dumoulin},
  {and} \bibinfo{person}{Aaron Courville}.} \bibinfo{year}{2018}\natexlab{}.
\newblock \showarticletitle{Film: Visual reasoning with a general conditioning
  layer}. In \bibinfo{booktitle}{\emph{AAAI Conference on Artificial
  Intelligence (AAAI)}}.
\newblock


\bibitem[\protect\citeauthoryear{Ravi and Larochelle}{Ravi and
  Larochelle}{2017}]%
        {ravi2016optimization}
\bibfield{author}{\bibinfo{person}{Sachin Ravi} {and} \bibinfo{person}{Hugo
  Larochelle}.} \bibinfo{year}{2017}\natexlab{}.
\newblock \showarticletitle{Optimization as a model for few-shot learning}.
\newblock  (\bibinfo{year}{2017}).
\newblock


\bibitem[\protect\citeauthoryear{Requeima, Gordon, Bronskill, Nowozin, and
  Turner}{Requeima et~al\mbox{.}}{2019}]%
        {requeima2019fast}
\bibfield{author}{\bibinfo{person}{James Requeima}, \bibinfo{person}{Jonathan
  Gordon}, \bibinfo{person}{John Bronskill}, \bibinfo{person}{Sebastian
  Nowozin}, {and} \bibinfo{person}{Richard~E Turner}.}
  \bibinfo{year}{2019}\natexlab{}.
\newblock \showarticletitle{Fast and flexible multi-task classification using
  conditional neural adaptive processes}. In \bibinfo{booktitle}{\emph{Annual
  Conference on Neural Information Processing Systems (NeurIPS)}}.
\newblock


\bibitem[\protect\citeauthoryear{Santoro, Bartunov, Botvinick, Wierstra, and
  Lillicrap}{Santoro et~al\mbox{.}}{[n.d.]}]%
        {santoro2016meta}
\bibfield{author}{\bibinfo{person}{Adam Santoro}, \bibinfo{person}{Sergey
  Bartunov}, \bibinfo{person}{Matthew Botvinick}, \bibinfo{person}{Daan
  Wierstra}, {and} \bibinfo{person}{Timothy Lillicrap}.}
  \bibinfo{year}{[n.d.]}\natexlab{}.
\newblock \showarticletitle{Meta-learning with memory-augmented neural
  networks}. In \bibinfo{booktitle}{\emph{International Conference on Machine
  Learning (ICML)}}.
\newblock


\bibitem[\protect\citeauthoryear{Snell, Swersky, and Zemel}{Snell
  et~al\mbox{.}}{2017}]%
        {snell2017prototypical}
\bibfield{author}{\bibinfo{person}{Jake Snell}, \bibinfo{person}{Kevin
  Swersky}, {and} \bibinfo{person}{Richard Zemel}.}
  \bibinfo{year}{2017}\natexlab{}.
\newblock \showarticletitle{Prototypical networks for few-shot learning}. In
  \bibinfo{booktitle}{\emph{Annual Conference on Neural Information Processing
  Systems (NeurIPS)}}.
\newblock


\bibitem[\protect\citeauthoryear{Vartak, Thiagarajan, Miranda, Bratman, and
  Larochelle}{Vartak et~al\mbox{.}}{2017}]%
        {vartak2017meta}
\bibfield{author}{\bibinfo{person}{Manasi Vartak}, \bibinfo{person}{Arvind
  Thiagarajan}, \bibinfo{person}{Conrado Miranda}, \bibinfo{person}{Jeshua
  Bratman}, {and} \bibinfo{person}{Hugo Larochelle}.}
  \bibinfo{year}{2017}\natexlab{}.
\newblock \showarticletitle{A meta-learning perspective on cold-start
  recommendations for items}. In \bibinfo{booktitle}{\emph{Annual Conference on
  Neural Information Processing Systems (NeurIPS)}}.
\newblock


\bibitem[\protect\citeauthoryear{Vinyals, Blundell, Lillicrap, Wierstra,
  et~al\mbox{.}}{Vinyals et~al\mbox{.}}{2016}]%
        {vinyals2016matching}
\bibfield{author}{\bibinfo{person}{Oriol Vinyals}, \bibinfo{person}{Charles
  Blundell}, \bibinfo{person}{Timothy Lillicrap}, \bibinfo{person}{Daan
  Wierstra}, {et~al\mbox{.}}} \bibinfo{year}{2016}\natexlab{}.
\newblock \showarticletitle{Matching networks for one shot learning}. In
  \bibinfo{booktitle}{\emph{Annual Conference on Neural Information Processing
  Systems (NeurIPS)}}.
\newblock


\bibitem[\protect\citeauthoryear{Volkovs, Yu, and Poutanen}{Volkovs
  et~al\mbox{.}}{2017}]%
        {volkovs2017dropoutnet}
\bibfield{author}{\bibinfo{person}{Maksims Volkovs}, \bibinfo{person}{Guangwei
  Yu}, {and} \bibinfo{person}{Tomi Poutanen}.} \bibinfo{year}{2017}\natexlab{}.
\newblock \showarticletitle{Dropoutnet: Addressing cold start in recommender
  systems}. In \bibinfo{booktitle}{\emph{Advances in Neural Information
  Processing Systems (NeurIPS)}}.
\newblock


\bibitem[\protect\citeauthoryear{Vuorio, Sun, Hu, and Lim}{Vuorio
  et~al\mbox{.}}{2019}]%
        {vuorio2019multimodal}
\bibfield{author}{\bibinfo{person}{Risto Vuorio}, \bibinfo{person}{Shao-Hua
  Sun}, \bibinfo{person}{Hexiang Hu}, {and} \bibinfo{person}{Joseph~J Lim}.}
  \bibinfo{year}{2019}\natexlab{}.
\newblock \showarticletitle{Multimodal Model-Agnostic Meta-Learning via
  Task-Aware Modulation}. In \bibinfo{booktitle}{\emph{Annual Conference on
  Neural Information Processing Systems (NeurIPS)}}.
\newblock


\bibitem[\protect\citeauthoryear{Wang and Blei}{Wang and Blei}{2011}]%
        {wang2011collaborative}
\bibfield{author}{\bibinfo{person}{Chong Wang} {and} \bibinfo{person}{David~M
  Blei}.} \bibinfo{year}{2011}\natexlab{}.
\newblock \showarticletitle{Collaborative topic modeling for recommending
  scientific articles}. In \bibinfo{booktitle}{\emph{ACM Knowledge Discovery
  and Data Mining (KDD)}}.
\newblock


\bibitem[\protect\citeauthoryear{Wang, Wang, and Yeung}{Wang
  et~al\mbox{.}}{2015}]%
        {wang2015collaborative}
\bibfield{author}{\bibinfo{person}{Hao Wang}, \bibinfo{person}{Naiyan Wang},
  {and} \bibinfo{person}{Dit-Yan Yeung}.} \bibinfo{year}{2015}\natexlab{}.
\newblock \showarticletitle{Collaborative deep learning for recommender
  systems}. In \bibinfo{booktitle}{\emph{ACM Knowledge Discovery and Data
  Mining (KDD)}}.
\newblock


\bibitem[\protect\citeauthoryear{Wang, Zhang, Zhao, Li, Xie, and Guo}{Wang
  et~al\mbox{.}}{2019}]%
        {wang2019multi}
\bibfield{author}{\bibinfo{person}{Hongwei Wang}, \bibinfo{person}{Fuzheng
  Zhang}, \bibinfo{person}{Miao Zhao}, \bibinfo{person}{Wenjie Li},
  \bibinfo{person}{Xing Xie}, {and} \bibinfo{person}{Minyi Guo}.}
  \bibinfo{year}{2019}\natexlab{}.
\newblock \showarticletitle{Multi-task feature learning for knowledge graph
  enhanced recommendation}. In \bibinfo{booktitle}{\emph{The World Wide Web
  Conference (WWW)}}.
\newblock


\bibitem[\protect\citeauthoryear{Xie, Girshick, and Farhadi}{Xie
  et~al\mbox{.}}{2016}]%
        {xie2016unsupervised}
\bibfield{author}{\bibinfo{person}{Junyuan Xie}, \bibinfo{person}{Ross
  Girshick}, {and} \bibinfo{person}{Ali Farhadi}.}
  \bibinfo{year}{2016}\natexlab{}.
\newblock \showarticletitle{Unsupervised deep embedding for clustering
  analysis}. In \bibinfo{booktitle}{\emph{International Conference on Machine
  Learning (ICML)}}.
\newblock


\bibitem[\protect\citeauthoryear{Xin and Jaakkola}{Xin and Jaakkola}{2014}]%
        {xin2014controlling}
\bibfield{author}{\bibinfo{person}{Yu Xin} {and} \bibinfo{person}{Tommi
  Jaakkola}.} \bibinfo{year}{2014}\natexlab{}.
\newblock \showarticletitle{Controlling privacy in recommender systems}. In
  \bibinfo{booktitle}{\emph{Annual Conference on Neural Information Processing
  Systems (NeurIPS)}}.
\newblock


\bibitem[\protect\citeauthoryear{Yao, Wei, Huang, and Li}{Yao
  et~al\mbox{.}}{2019}]%
        {yao2019hierarchically}
\bibfield{author}{\bibinfo{person}{Huaxiu Yao}, \bibinfo{person}{Ying Wei},
  \bibinfo{person}{Junzhou Huang}, {and} \bibinfo{person}{Zhenhui Li}.}
  \bibinfo{year}{2019}\natexlab{}.
\newblock \showarticletitle{Hierarchically structured meta-learning}. In
  \bibinfo{booktitle}{\emph{International Conference on Machine Learning
  (ICML)}}.
\newblock


\bibitem[\protect\citeauthoryear{Zhang, Yao, Sun, and Tay}{Zhang
  et~al\mbox{.}}{2019}]%
        {zhang2019deep}
\bibfield{author}{\bibinfo{person}{Shuai Zhang}, \bibinfo{person}{Lina Yao},
  \bibinfo{person}{Aixin Sun}, {and} \bibinfo{person}{Yi Tay}.}
  \bibinfo{year}{2019}\natexlab{}.
\newblock \showarticletitle{Deep learning based recommender system: A survey
  and new perspectives}.
\newblock \bibinfo{journal}{\emph{ACM Computing Surveys (CSUR)}}
  (\bibinfo{year}{2019}).
\newblock


\bibitem[\protect\citeauthoryear{Zheng, Tang, Ding, and Zhou}{Zheng
  et~al\mbox{.}}{2016}]%
        {zheng2016neural}
\bibfield{author}{\bibinfo{person}{Yin Zheng}, \bibinfo{person}{Bangsheng
  Tang}, \bibinfo{person}{Wenkui Ding}, {and} \bibinfo{person}{Hanning Zhou}.}
  \bibinfo{year}{2016}\natexlab{}.
\newblock \showarticletitle{A neural autoregressive approach to collaborative
  filtering}. In \bibinfo{booktitle}{\emph{International Conference on Machine
  Learning (ICML)}}.
\newblock


\end{thebibliography}


\end{document}